\def\year{2019}\relax
\documentclass[letterpaper]{article} 
\usepackage{aaai19}  
\usepackage{times}  
\usepackage{helvet} 
\usepackage{courier}  
\usepackage[hyphens]{url}  
\usepackage{graphicx} 
\usepackage{graphicx}  
\usepackage{amsmath}
\usepackage{booktabs}
\usepackage{colortbl}
\usepackage{enumitem}
\usepackage{tabularx}
\usepackage{xcolor}

\title{The Language of Dialogue Is Complex}

\author{Alexander Robertson,\textsuperscript{\rm 1} Luca Maria Aiello,\textsuperscript{\rm 2} Daniele Quercia\textsuperscript{\rm 2}\\
\textsuperscript{\rm 1}University of Edinburgh,
\textsuperscript{\rm 2}Nokia Bell Labs\\
alexander.robertson@ed.ac.uk, luca.aiello@nokia-bell-labs.com, daniele.quercia@nokia-bell-labs.com
}

\begin{document}

\maketitle

\begin{abstract}
Integrative Complexity (IC) is a psychometric that measures the ability of a person to recognize multiple perspectives and connect them, thus identifying paths for conflict resolution. IC has been linked to a wide variety of political, social and personal outcomes but evaluating it is a time-consuming process requiring skilled professionals to manually score texts, a fact which accounts for the limited exploration of IC at scale on social media. We combine natural language processing and machine learning to train an IC classification model that achieves state-of-the-art performance on unseen data and more closely adheres to the established structure of the IC coding process than previous automated approaches. When applied to the content of 400k+ comments from online fora about depression and knowledge exchange, our model was capable of replicating key findings of prior work, thus providing the first example of using IC tools for large-scale social media analytics.
\end{abstract}

\section{Introduction}

Integrative complexity (IC) is a psychometric that measures the degree to which a person has engaged in two cognitive processes, given a particular topic or issue~\cite{sts1992}. The first is differentiation: the recognition of multiple perspectives for the issue at hand. The second is integration: having identified such perspectives, the person demonstrates how these perspectives are connected. The lowest end of the IC spectrum is associated with inflexible, fixed perspective thinking and the highest end with integrating groups of perspectives in an elaborate, hierarchical fashion. Table~\ref{ic_description} outlines the seven IC bands as described by \citeauthor{baker1990coding}~\shortcite{baker1990coding}.

\begin{table}[ht!]
\centering
\resizebox{0.95\columnwidth}{!}{%
\begin{tabular}{cccl}
\toprule
IC & Differentiation & Integration & Details \\
\midrule
1 & None & None & No evidence of IC. \\
2 & Emergent & None & Some acknowledgment of differing \\
 & & & views \\
3 & Explicit & None & At least two perspectives stated. \\
4 & Explicit & Emergent & Connections suggested, not stated. \\
5 & Explicit & Explicit & All perspectives connected in a new \\
& & & perspective. \\
6 & Explicit & Explicit & High level of integration \\
7 & Explicit & Explicit & Overarching perspective, detailing \\
& & & relationship between alternatives. \\
\bottomrule
\end{tabular}%
}
\caption{Description of the seven levels of integrative complexity and the degree to which they exhibit evidence of cognitive differentiation and integration.}
\label{ic_description}
\end{table}

IC has been applied to a wide range of source materials, including diplomatic communications, political speeches, personal correspondence and legal judgments~\cite{sts1992}. As a result, it has been presented as a powerful predictor for a variety of outcomes, such as whether an international crisis will end in conflict~\cite{suedfeld1977integrative,suedfeld1977war} or how far along in a term a president is \cite{thoemmes2007integrative}. In addition, IC levels have been linked with many other factors, such as aggression~\cite{winter1993slot} and political preferences~\cite{conway2016conservatives}. Despite varied and interesting findings, IC is argued to be under-utilized in research~\cite{conway2014automated}. This is attributed to the time-consuming nature of manual scoring of texts. Furthermore, becoming qualified to determine IC requires several weeks of intensive training. Remedying these issues should therefore see IC used more often and at much larger scales. An attractive proposition, it has motivated the development of automated approaches to IC scoring. This is in spite of the perceived difficulty of the task by experts: while they state that an automated system able to perform IC scoring would be a major advance, they simultaneously warn that IC ``does not rely on simple content-counting rules'' and ``cannot be reduced to a simple [...] content analysis system''~\cite{baker1990coding}. 

This has not discouraged attempts. Following a brief description of IC, we describe these prior attempts (\S\ref{section:prior_work}) and we then present our contributions:
\begin{itemize}
	\item We build and make publicly available a machine learning model for automated scoring of IC which uses syntactic features that are theoretically well-motivated by the IC framework (\S\ref{sec:method});
	\item We test our model on the official IC scoring test and achieve state-of-the-art results, with a F1 score of almost 25\% higher than previous approaches (\S\ref{sec:results});
	\item We conduct for the first time an analysis of IC at scale by applying our tool to over 400k textual snippets from Reddit. Results obtained on a support-based forum focused on mental health match theoretical expectations, thus providing initial evidence of external validity to our tool  (\S\ref{sec:social}) and setting the stage for its usage in the context of large-scale social media analytics (\S\ref{sec:conclusion}).
\end{itemize}

\section{Prior work}\label{section:prior_work}

\subsection{Linguistic style analysis}\label{section:prior_work:nlp}

The study of linguistic style in text and conversations has been related to a number of outcomes.

On Twitter, researchers have investigated the use of specific markers that are predictive of the initiation of a conversation~\cite{boyd2010tweet} and found that linguistic affinity between participants fosters continued engagement~\cite{budak2013participation}. From text, one can also predict more intangible properties associated to verbal expressions. Multimodal features of online threads can predict the perceived interestingness of the themes discussed~\cite{de2009makes}. The combined use of topic detection and sentiment analysis on Twitter has been used  to extract higher-level emotional properties such sympathy, apology, and complaint~\cite{kim2012you}. Statistical stylometry has been used to evaluate the quality of literary writing and to identify successful pieces of literature~\cite{ashok2013success}.

Conversation style has also implications on the social processes involving participants. The evolution of discussion topics over time unearths patterns of social identity and cohesion~\cite{purohit2014understanding}. The language complexity, and emotions expressed in a conversation~\cite{danescu2012echoes,tchokni2014emoticons} echoes the power differential of participants. A considerable amount of work has been done to understand the connection between linguistic style and conflict. Linguistic cues such as markers of agreement and confidence distinguish between productive and unproductive discussions~\cite{niculae2016conversational}. Rhetorical prompts deployed in the very first conversation exchanges are predictive of emergence of conflict~\cite{zhang18conversations}. Antisocial behavior is also impacted by the mood of the context surrounding the discussion~\cite{cheng2017anyone} and exacerbated by individuals who attempt to steer the discourse towards irrelevant topics~\cite{cheng2015antisocial}.

In this work, instead of studying linguistic markers that create conflict, we focus on how language can bring reconciliation and peace by recognizing different points of view and integrating them.

\subsection{Automated IC methods}\label{section:prior_work:autoic}

The success of prior work runs counter not only to the technological predictions of experts, but also to the theory that underpins IC. Specifically, that it is a measure of \emph{structure} rather than \emph{content}. Integrative Complexity, as scored by skilled humans, is concerned not with \emph{what} we say, but \emph{how} we say it. The two extant automatic coding methods, detailed below, are both focused on content.

\citeauthor{ambili2014automated}~\shortcite{ambili2014automated} trained models to predict low, medium or high IC using as features the text length, the vocabulary used, and a metric based on the semantic coherence of the text~\cite{li2003approach}. Using the words in the first sentence of a text, pairwise comparisons with all subsequent words was performed on the basis of their connection in the WordNet~\cite{wordnet} knowledge-base. Specifically, the normalized product of the minimum path length between two words and the depth in the knowledge-base hierarchy of their lowest common hypernym. This value, along with text length and one-hot encoding of the words in the text, were used as features in a variety of classifier models. The system achieved an F1 score of around 0.8.

\citeauthor{conway2014automated}~\shortcite{conway2014automated} created a rule-based system based on the presence of specific vocabulary items thought to be indicative of differentiation or integration. The count of each word is then weighted by custom values. If the differentiation keywords do not reach a threshold of 3, then no integration keywords are considered in calculating the final score. Because weights are real-valued, the system does not output classes but real numbers between 1 and 7. The system was evaluated using correlation with human scorers, as well as by attempting to replicate prior outcomes on small human-coded datasets.

\begin{table*}[t!]
\centering
\resizebox{1.99\columnwidth}{!}{%
\begin{tabular}{@{}rccccl@{}}
\toprule
Material & Source & Texts & Tokens/text & Usage & Score distribution \\
\midrule
Official IC Practice Sets & Suedfeld, 1992 & 156  & $\mu$ 57.5 ($\sigma$ 41.7) & Training   & \includegraphics[width=0.30\columnwidth]{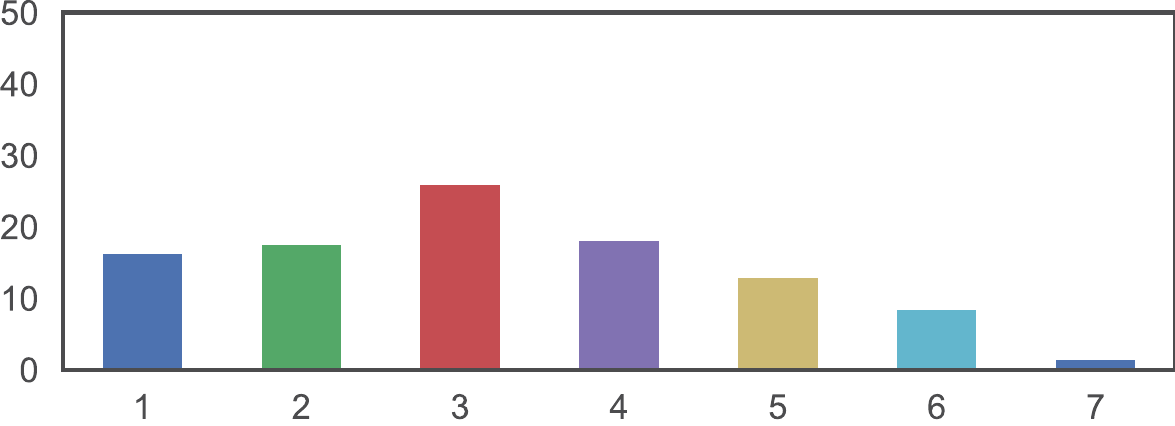}\\
Heritability & Conway, 2014	 & 310  & $\mu$ 92.7 ($\sigma$ 52.4) & Training & \includegraphics[width=0.30\columnwidth]{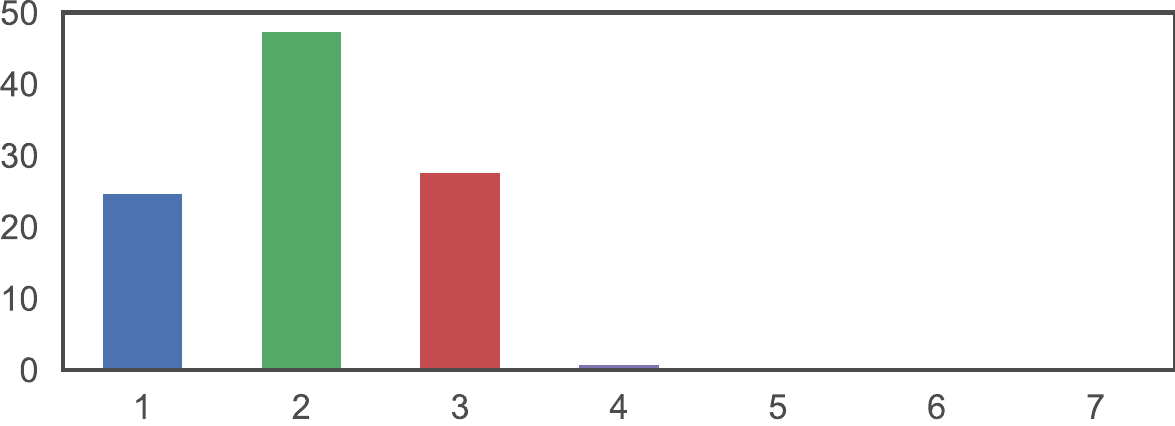}\\
Early Christian Writings & Conway, 2014	 & 173  & $\mu$ 117.6 ($\sigma$ 68.1) & Training & \includegraphics[width=0.30\columnwidth]{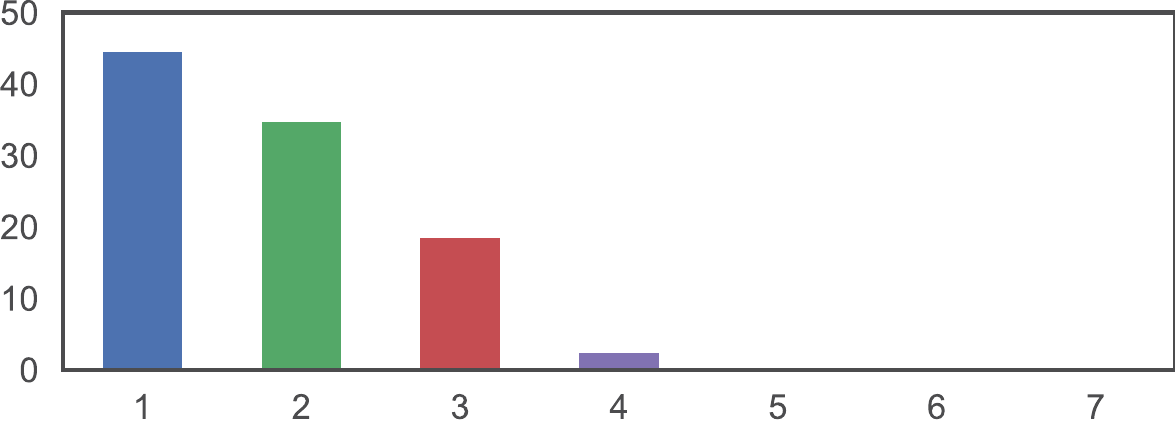}\\
Official IC Coding Test & Suedfeld, 1992 & 30   & $\mu$ 72.7 ($\sigma$ 30.7) & Evaluation & \includegraphics[width=0.30\columnwidth]{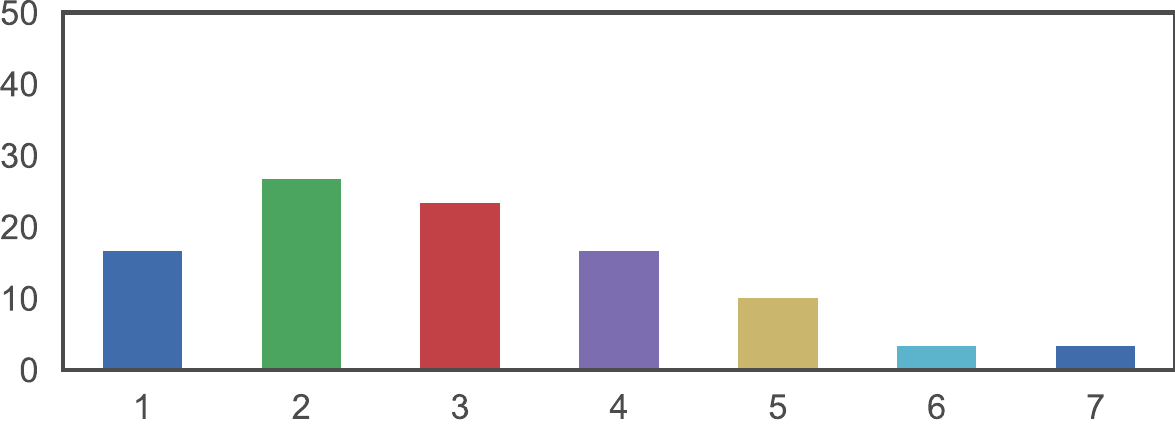}\\
\bottomrule
\end{tabular}%
}
\caption{Datasets used in experiments, and their properties.}
\label{datasets}
\end{table*}

\subsection{Challenges}
Each of these approaches adjusts the structure and assumptions of the IC system. \citeauthor{ambili2014automated} reduce the number of classes from seven to three bins: low (1,2), medium (3,4,5) and high (6,7). Although this reduction in task difficulty was motivated by lack of training data, the bins are not well aligned with the structure of the IC bands they represent. A band of 1 is not simply ``low'' but a complete absence of IC, while bands 3 to 5 represent very different levels of differentiation and integration.

Similarly, the real-valued output of \citeauthor{conway2014automated} does not strictly align with the bands of the IC scoring system. Each band is a label for the absence or presence of particular properties and there is no provision in the literature for non-integer scores~\cite{baker1990coding}. It is not appropriate to use linear correlation methods to compare these pseudo-continuous variables to categorical ones, since this may result in higher correlations than when the real-valued predictions are rounded to become integer values.

The seven IC bands in Table~\ref{ic_description} are most properly treated as ordinal variables. Differences between bands cannot be numerically quantified. It is not possible to claim that the difference between an IC band of 2 and 3 is equivalent to that between 3 and 4. This enables us to cast the scoring task as a classification problem, with seven distinct class labels.

Last, and most importantly, both previous approaches focus on semantics and ignore syntax (\emph{how} people express concepts). In the following, we will show that syntactic information is crucial to generalize automated scoring to text whose nature is very different from the text upon which the model training is performed.

\section{Methodology}\label{sec:method}

\subsection{Training data}

For training and evaluation, we use the datasets outlined in Table~\ref{datasets}. We follow \citeauthor{conway2014automated} in using some sources as training material, while leaving others aside for evaluation. The Official Practice Sets and Official Coding Test are taken from the Suedfeld's Electronic Complexity Workshop. The rest were kindly supplied by Conway: the Heritability texts cover student responses to prompts on a range of topics (religion, death, assertiveness); Early Christian Writings are randomly sampled from the New Testament. All IC scoring was performed by trained humans, with texts being scored by multiple scorers. The ``unofficial'' materials are somewhat longer and more numerous, but also show a distinct lack of variety in the range of IC bands represented. Where there was slight disagreement between scorers (the IC handbook permits a difference of up to 2) for a particular text, the average was taken---we have therefore rounded these scores to the nearest whole integer. Examples of texts belonging to the 7 IC bands are reported in Table~\ref{tab:ic_example_text}, in the Appendix.

\subsection{Features sets}

The automated techniques described in \S\ref{section:prior_work:autoic} take a semantic approach. Semantic features include actual words or phrases, information about senses of words or phrases, or information about classes of words in terms of their meaning (e.g., whether they are positive or related to a particular topic). An alternative, closer to \emph{how} than \emph{what}, is a syntactic approach. We therefore distinguish two types of features along these lines. Syntactic features include more abstract properties, related to the way language is meaningfully structured. For example, the syntactic role played by a word within a text (e.g., noun, adjective) or the syntactic relations between words (e.g., direct object of a verb). Below, we detail both the semantic features (vocabulary-based) and the two families of syntactic features (POS tags and dependency subtrees) we considered. We extract syntactic features using the CNN-based tagger provided by the spaCy python package, which has an accuracy of 97\% for POS tagging and 90\% for dependency labeling on English texts.

\vspace{4pt}\noindent\textbf{Vocabulary.} Following the approach of \citeauthor{conway2014automated}, we use the IC handbook \cite{baker1990coding} to identify key phrases which are said to be associated with each particular IC band. The original set of words includes some ``content-insensitive'' words (adverbs like \textit{however} or \textit{yet}) and some words that directly refer to differentiation and integration processes (e.g., \textit{compromise}, \textit{compensate}, \textit{reconciliation}). We expand this list with synonyms and related terms by searching for each key phrase in the ConceptNet knowledge-base \cite{conceptnet}. We clean the resulting expanded list by manually filtering out items not likely to be linked to differentiation/integration. This is a binary feature, representing the presence or absence of a given vocabulary item. Rather than creating a feature for each form a word or phrase can take, we lemmatize all keywords and search for these in a lemmatized version of the input text. Text length has been shown to be vacuously predictive of IC~\cite{baker1990coding}, so we use binary features rather than counts or weighted counts as in \citeauthor{conway2014automated} \shortcite{conway2014automated}---to lower the risk of implicitly encoding the length of the the text as a feature. Finally, we compute two binary features, indicating the presence of vocabulary related to differentiation or integration. We extract 312 semantic features.

\vspace{4pt}\noindent\textbf{POS tags.} Using the Penn Treebank tag set, tokens in a text are labelled according to their syntactic part of speech (e.g., as particular types of nouns, adjectives or verbs). Counts of these tags are then normalized by the total number of words in the text. We extract 45 syntactic features.

\vspace{4pt}\noindent\textbf{Dependency subtrees.} A dependency parser labels the relationship between words. This relationship shows which words modify others: if word A is modified by word B, then A is a child of B. These relations form a graph. By starting at each node in the graph and determining its descendants, subtrees can be extracted. In these subtrees, edges between words are labeled with the type of relationship between the two words. For example, the sentence ``the cat sleeps'' is converted to $sleeps \xrightarrow{\text{nsubj}} cat \xrightarrow{\text{det}} the$. The article \textit{the} is the determiner (det) of the noun \textit{cat} and \textit{cat} is the nominal subject (nsubj) for the verb \emph{sleeps}. The resulting features encode only the labels on the edges, not the labels in the nodes. In the example, the feature extracted are \textit{nsubj}, \textit{det} (subtrees of length 1) and \textit{nsubj\_det} (subtree of length 2). Similar to vocabulary features, subtree features are binary, set to 1 if the subtree is detected at least once in the text. Due to the small size of our training datasets and having the goal of extracting complex syntactic structures, we extract subtrees with a number of edges up to 5. Prior work by \citeauthor{vosoughi2016tweet} \shortcite{vosoughi2016tweet} used a maximum length of 2, but were working with tweets---not only a much larger dataset, but also much smaller individual texts. We keep as features only those subtrees which appear in the training datasets with a frequency of at least 5. Counts of subtrees are normalized by the total number of subtrees extracted. We extract 280 syntactic features.

\vspace{4pt}\noindent\textbf{LIWC.} LIWC is a standard dictionary of 2,300 English words grouped in 72 categories \cite{tausczik2010psychological}. These categories are generally abstract and aim at capturing markers of emotional and psychological expressions that do not hinge on the particular topic of the text. Examples of categories include expressions focused on the future or verbs referred to the human perceptual sphere. Each word may belong to multiple categories. Similar to the vocabulary features, we use 72 binary scores rather than counts: each feature is set to 1, if at least a word in the respective LIWC category is present in the text.

\section{Classification results}\label{sec:results}

\subsection{Experimental setup}

\noindent\textbf{Classifier.} We evaluate each feature set by training an ensemble of decision trees with gradient boosting~\cite{xgboost}, implemented by the xgboost package. This model is well-suited to the small datasets we are working with, makes it easy to interpret the contribution of individual features, and is able to ignore any vacuous features that may be present. This last property is useful to avoid overfitting since some feature sets are large.

\vspace{4pt}\noindent\textbf{Baselines.} We compare our classifier against four baselines. First, we include a simple baseline that always predicts the most common class. Second, we consider a method that uses text length, measured as the number of words, as the only feature. Third, we use text sentiment. Among the many sentiment analysis tools available, we chose Vader \cite{gilbert2014vader}, a state-of-the-art sentiment analysis technique widely used on noisy text. Vader outputs a score from -1 (most negative) to 1 (most positive). Last, we compare our results against the AutoIC tool \cite{conway2014automated}. AutoIC  has been previously trained by Conway on the same data we use, and that makes it an appropriate choice for a fair comparison. AutoIC outputs real numbers rather than integers. However, IC bands as described in the IC handbook \cite{baker1990coding} represent whether particular properties are present/absent, rather than having a continuous value. Failing to reach a threshold value is therefore best considered as evidence of failing to detect sufficient evidence for that threshold. We therefore we apply the floor function to AutoIC's output.

\vspace{4pt}\noindent\textbf{Evaluation metrics.} We compare different approaches and feature sets using the F1 score. For each of the 7 classes, F1 score is calculated. The harmonic mean is then weighted by the total number of examples of that class, with the overall average reported. Casting IC scoring as a discrete classification problem is justified by the official IC handbook, which makes no provision for interpreting IC as a continuous variable. Additionally, we provide confusion matrices for each feature set to broadly illustrate the strengths and weaknesses of each with regard to the seven IC classes. Finally, in order to measure how close a model's predictions are to the true class labels we calculate the Mean Squared Error (MSE) between the two. This is possible because the class labels are ordinal, even if not continuous, and have a sensible ordering. However, we acknowledge that the magnitude of the difference between adjacent class labels is not likely to be linear. Misclassifying an IC text as 3 instead of 2, where the difference is based purely on degree of differentiation, should not be penalized to the same degree as misclassifying as 4 instead of 3, where the difference is now due to integration.

\subsection{Cross-validation on training data}

\begin{figure}[t]
		\includegraphics[width=0.99\columnwidth]{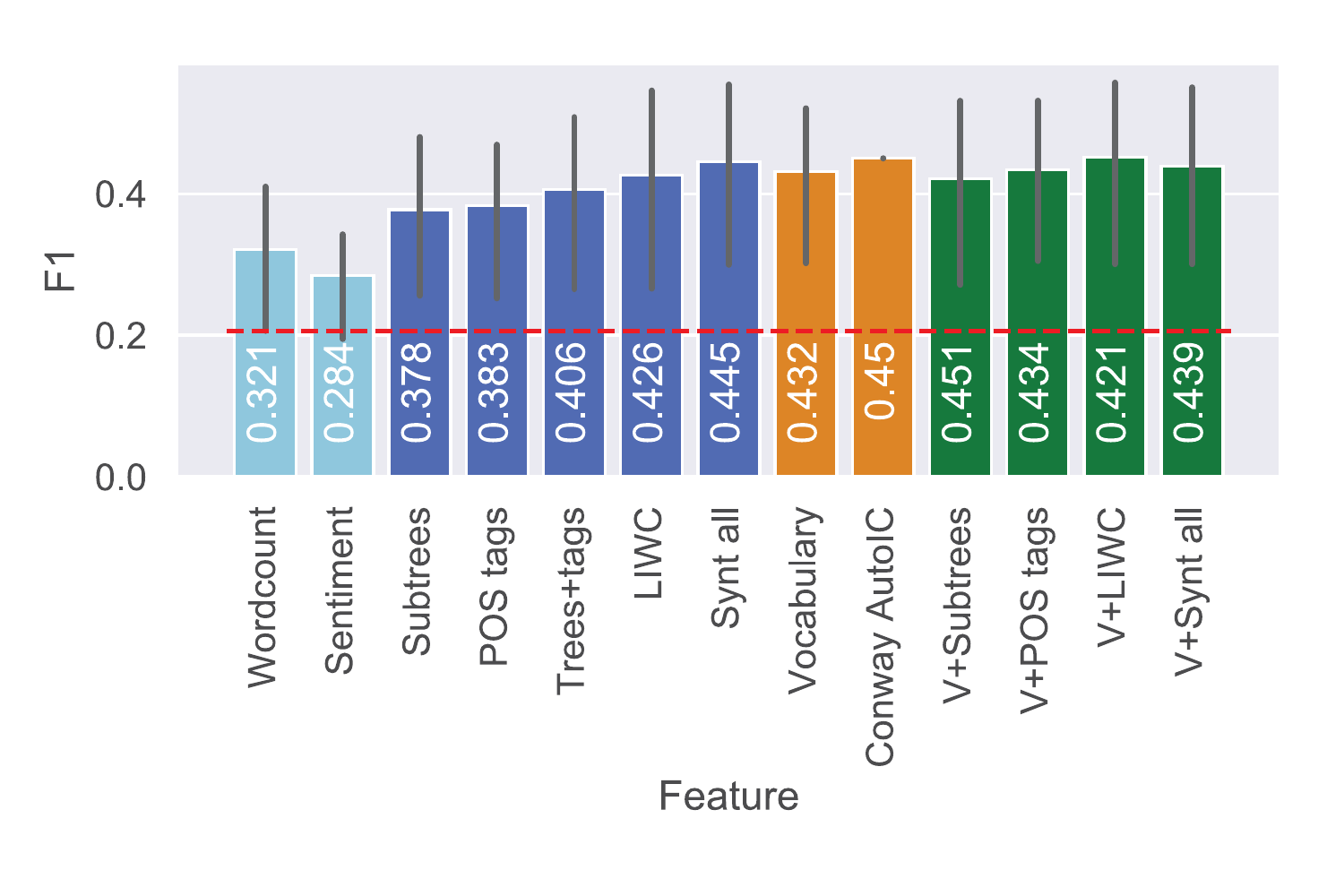}
\caption{Classifier performance on the training dataset on 5 cross-validation folds, in terms of F1 score. Error bars show variance across folds. Wordcount and sentiment baselines are in light blue, syntactic features in blue, semantic in orange, joint in green. The red line shows the baseline result for the most frequent class predictor.}
\label{feature_results}
\end{figure}

\begin{figure}[t]
\includegraphics[width=0.95\columnwidth]{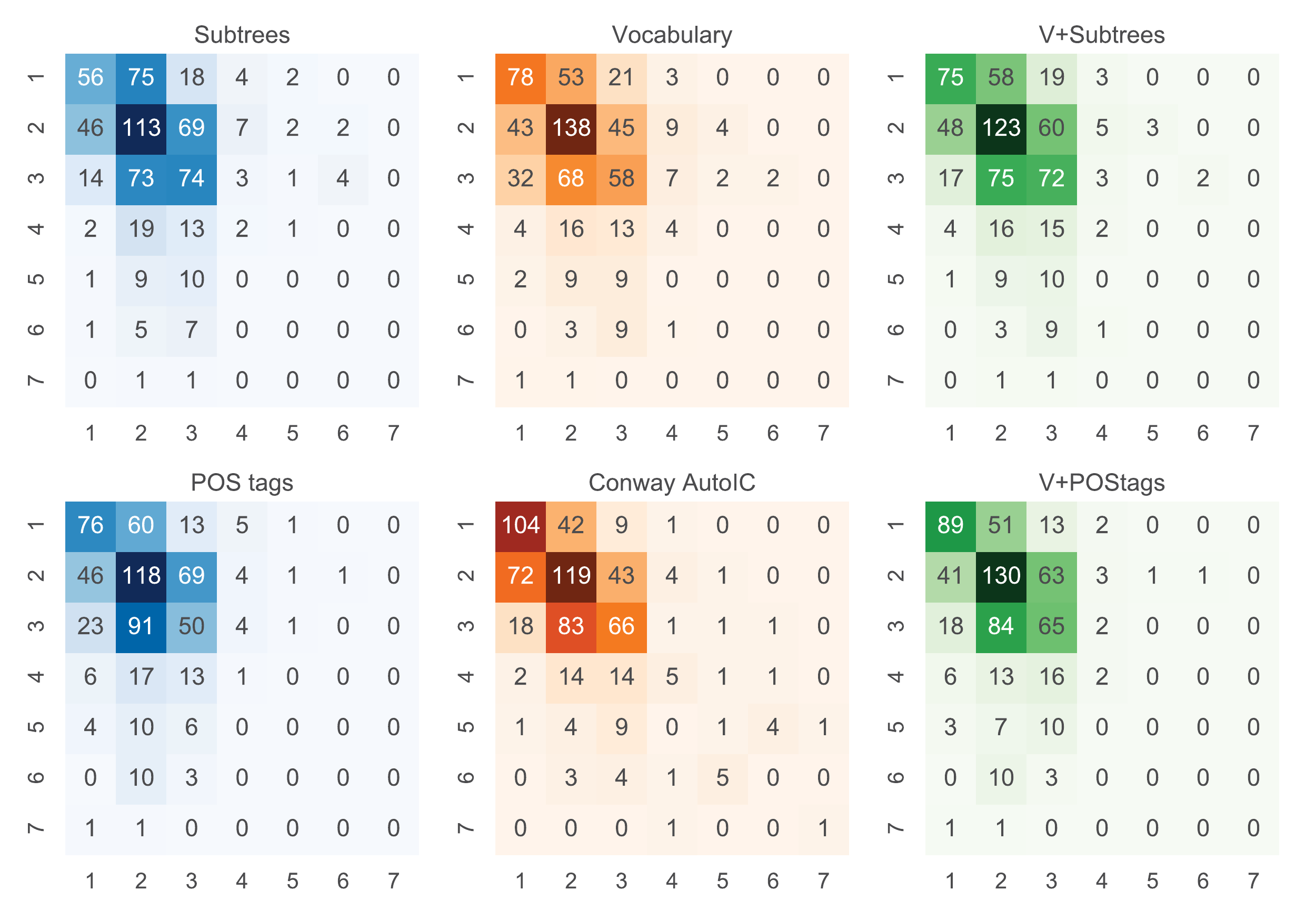}
\caption{Confusion matrices for each classifier trained on syntactic (blue), semantic (orange) or joint (green) features, evaluated on the training datasets summed across cross-validation folds. Rows are the true classes, columns are classifiers' predicted classes.}
\label{confmat_train}
\end{figure}

In the first experiment, we use 5-fold cross-validation on the training set and report the mean F1 score (with standard deviation) achieved across all folds, for each feature set individually and some of them in combination, along with the baselines. Figure~\ref{feature_results} shows the mean F1 score over all five cross-validation folds. 
The sentiment and word-count approaches are the worst-performing ones after the naive baseline. When combined with other feature sets, they never yield any performance improvement. Syntactic features perform better (LIWC especially) and work also well in combination, being able to achieve a F1 of .445 when put together. In this case, semantic approaches perform roughly as well as syntactic features combined. Merging syntactic and semantic feautures yields no improvement on the results.
The imbalanced dataset is problematic: few models can correctly predict IC above 4. The only notable exception is Vocabulary, which manages to identify 4 (out of 37) such documents. AutoIC fares better: 5/37 for band 4, 1/20 for band 5, 1/2 for band 7 (Figure~\ref{confmat_train}). This may be attributed to the method used by AutoIC to weight vocabulary items in terms of their contribute to differentiation/integration, compared to our binary approach. AutoIC may also have a broader list of keywords which are coincidentally present in the higher IC texts. The mean MSE values across all cross-validation folds for selected feature subsets are shown in the first row of Table~\ref{rmse_all}. AutoIC's lower error is due to the fact that their model predicts a wider range of bands, whilst still failing to classify them correctly. In general, model predictions are within a reasonably tight range of the true labels.

\subsection{Prediction on heldout data}

In the second experiment, we retain all the data used in the first experiment for training and tuning. We fine-tune model parameters through cross-validated grid search. A large ensemble of relatively deep trees (500, with a maximum depth of 6), with 80\% subsampling, gave the best performance. Subsampling sets the proportion of training data, selected at random from all available data, that each decision tree receives during training and helps to reduce the effect of overfitting a model to the data. We then train a new model, one per feature set, with the best parameters. These models are evaluated using the official IC coding test (last row in Table~\ref{datasets}). Crucially, the creators of AutoIC did not use the official coding test to manually identify any differentiation/integration words or phrases, making this an especially fair comparison of systems.

\begin{table}[t!]
\centering
\resizebox{0.95\columnwidth}{!}{%
\begin{tabular}{@{}lcccccc@{}}
\toprule
           &  AutoIC & Subtrees	& POS tags  & Vocab.  & V+Subtrees  & V+POStags         \\ \midrule
Cross-val. & 1.08    & 1.63 		& 1.70 		 	& 1.59		& 1.43 		& 1.48		 	\\
Heldout    & 1.33    & 2.87 		& 2.13 			& 3.40 	  & 2.60		& 1.80 		 \\ \bottomrule
\end{tabular}%
}
\caption{Mean squared error for predictions made in the cross-validation and heldout experiments. Result for Conway's AutoIC is shown for reference.}
\label{rmse_all}
\end{table}

\begin{table*}[t!]
\centering
\begin{tabular}{@{}ccccccccccc@{}}
\toprule
\multicolumn{1}{l}{}        &         & \multicolumn{3}{c}{V+POStags}     & \multicolumn{3}{c}{Conway AutoIC} & \multicolumn{3}{c}{Difference}                                                                \\ \cmidrule{3-11}
\multicolumn{1}{r}{IC band} & Support & Precision    & Recall    & F1     & Precision    & Recall    & F1     & \multicolumn{1}{l}{Precision} & \multicolumn{1}{l}{Recall}    & \multicolumn{1}{l}{F1}        \\ \midrule
1                           & 5       & 0.71         & 1.00      & 0.83   & 0.57         & 0.80      & 0.67   & +0.14 & +0.20 & +0.16   \\
2                           & 8       & 0.45         & 0.62      & 0.53   & 0.45         & 0.62      & 0.53   & +0.00    & +0.00    & +0.00    \\
3                           & 7       & 0.33         & 0.43      & 0.38   & 0.40         & 0.29      & 0.33   & -0.07 & +0.14 & +0.05 \\
4                           & 5       & 1.00         & 0.6       & 0.75   & 0.5          & 0.40      & 0.44   & +0.50 & +0.20 & +0.31 \\
5                           & 3       & 0.00         & 0.00      & 0.00   & 0.00         & 0.00      & 0.00   & 0.00  & 0.00  & 0.00  \\
6                           & 1       & 0.00         & 0.00      & 0.00   & 0.00         & 0.00      & 0.00   & 0.00  & 0.00  & 0.00  \\
7                           & 1       & 0.00         & 0.00      & 0.00   & 0.00         & 0.00      & 0.00   & 0.00  & 0.00  & 0.00  \\ \midrule
\multicolumn{2}{r}{Average}     & 0.48         & 0.53      & 0.49   & 0.39         & 0.43      & 0.40   & +0.09 & +0.13 & +0.09 	  \\ \bottomrule
\end{tabular}
\caption{Classification report for the best-performing model, V+POStags, on the official 30 item IC coding test. Results of Conway's AutoIC tool are provided for reference.}
\label{classification_report}
\end{table*}

\begin{figure}[ht]
\includegraphics[width=0.95\columnwidth]{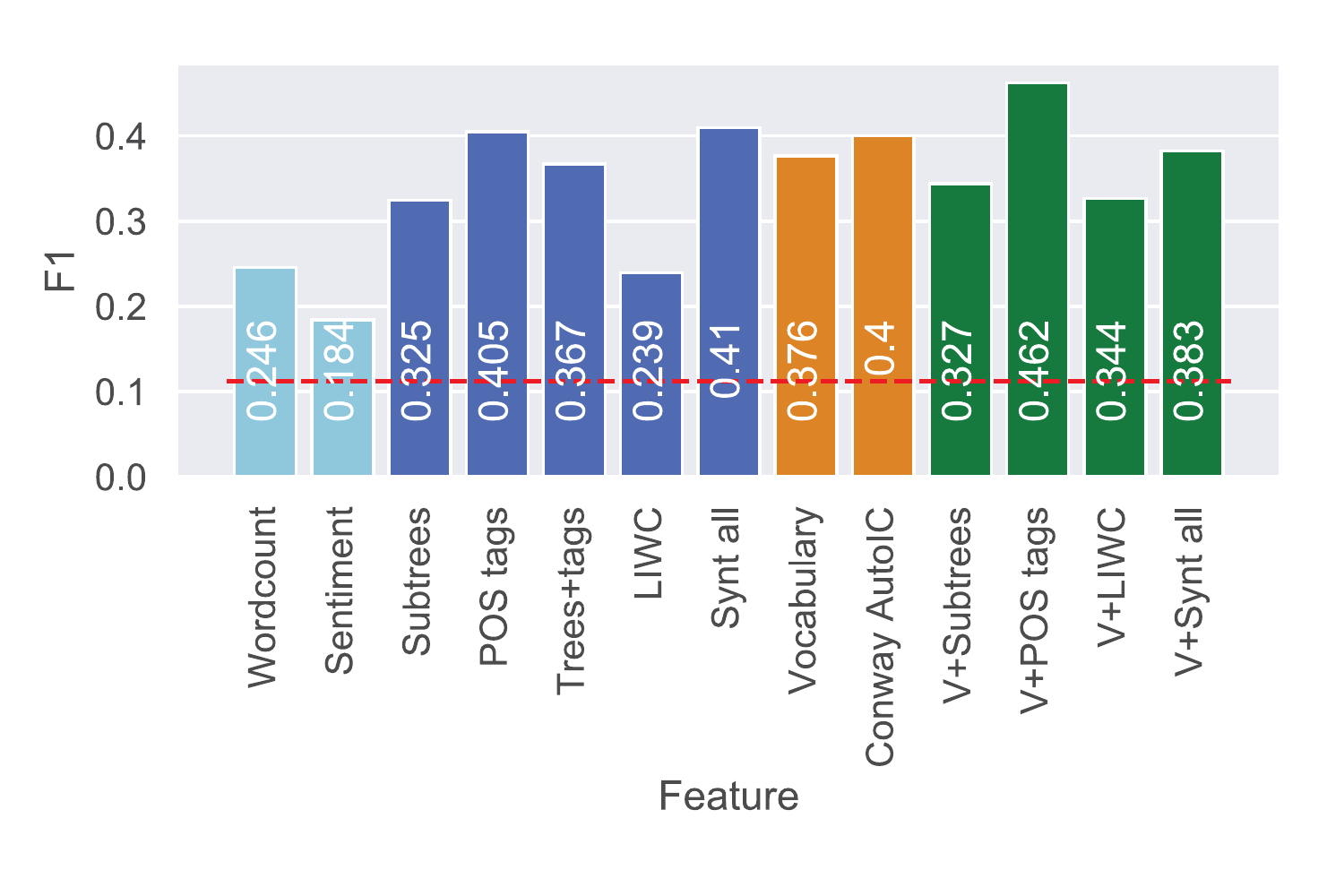}
\caption{Classifier performance on the heldout dataset, measured by F1 score. Word count and sentiment baselines are in light blue, syntactic features in blue, semantic in orange, joint in green. The red line shows the populous baseline.}
\label{feature_results_heldout}
\end{figure}

Figure~\ref{feature_results_heldout} shows the F1 score on the official IC scoring test. Unlike in the first prediction experiment, some syntax-based models---in particular, POS Tags and all syntactic features combined---outperform vocabulary features and are comparable to AutoIC. LIWC's F1 drops considerably compared to the cross-validation setting. The best-performing model uses both semantic and syntactic features to achieve an F1 score of 0.462. This is a non-negligible improvement over AutoIC, which scores an F1 of 0.400.

\begin{figure}[ht]
\includegraphics[width=0.95\columnwidth]{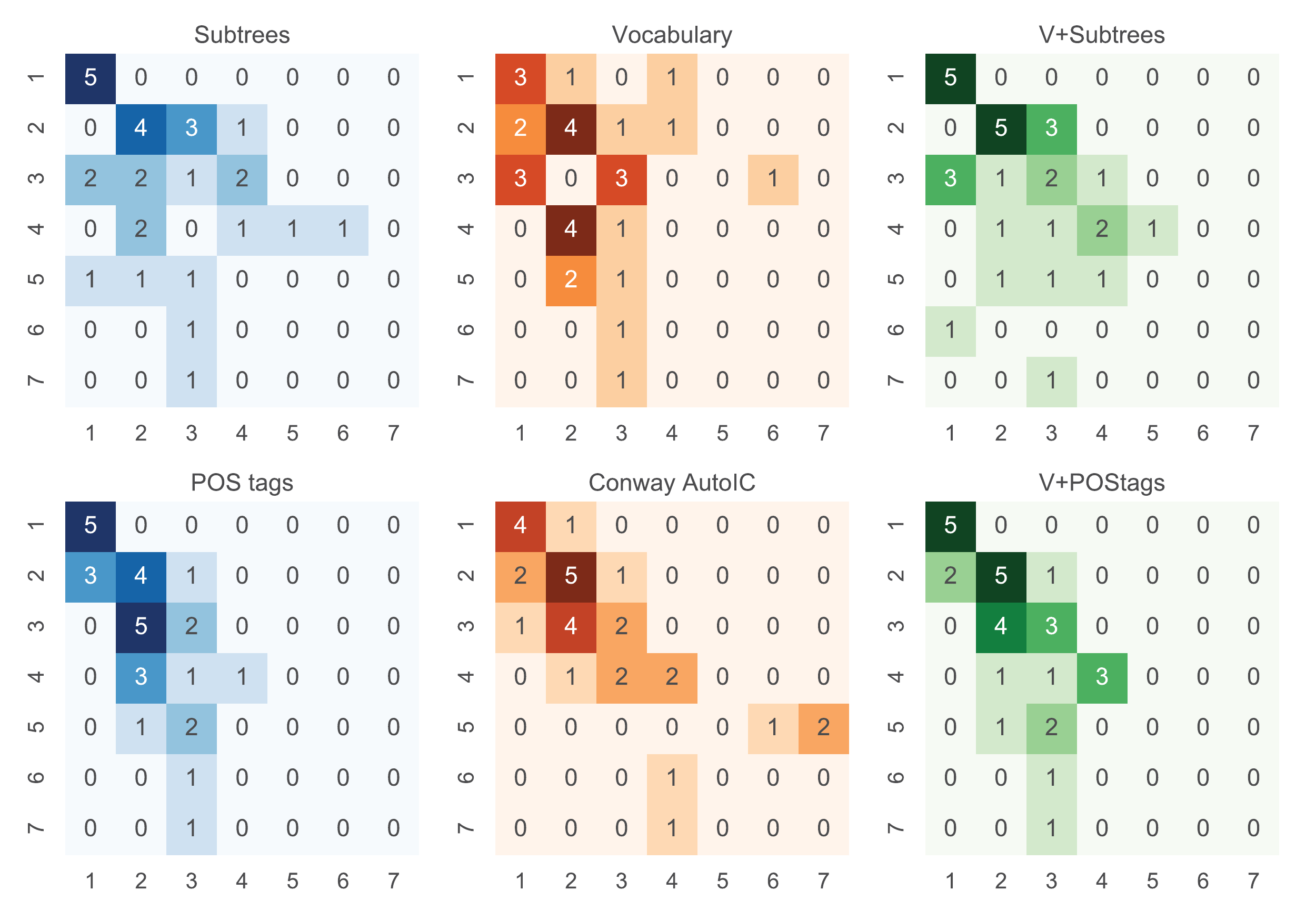}
\caption{Confusion matrices for each classifier trained on syntactic (blue), semantic (orange) or joint (green) features, evaluated on the heldout dataset, summed across cross-validation folds. Rows are the true classes, columns are classifiers' predicted classes.}
\label{confmat_heldout}
\end{figure}

The confusion matrices in Figure~\ref{confmat_heldout} show that while most models correctly classify IC band 1, no model correctly predicts an IC band higher than 4. Also, dependency Subtrees are the only features which result in a model producing many predictions above IC band 4, though not with any accuracy. MSE is shown in Table~\ref{rmse_all}. Again, Conway's AutoIC makes predictions which are numerically closer to the real class labels (again, due to the wider range of incorrect class labels predicted) but this does not translate into better F1 scores. This outcome highlights the importance of reporting multiple relevant metrics in order to show a more nuanced picture of model performance.

For the best-performing model, a classification report is shown in Table~\ref{classification_report}. Performance is generally high for classes 1 to 4, but no predictions above these bands are made. This is not especially surprising, since these higher bands are generally very rare and appear mainly in the official training materials for IC scoring. When compared to Conway's AutoIC, a similar situation for high IC bands emerges. In general, however, the combination of semantic and syntactic features significantly outperforms the purely semantic AutoIC system.
	
\begin{table*}[t]
\centering
\resizebox{1.9\columnwidth}{!}{%
	\begin{tabular}{@{}rccrccrcc@{}}
	\toprule
	\multicolumn{3}{c}{IC = 1}                            & \multicolumn{3}{c}{IC = 2}                            & \multicolumn{3}{c}{IC = 3}                            \\ \midrule
	Feature                       & Value & Contribution & Feature                       & Value & Contribution & Feature                       & Value & Contribution \\ \midrule
	has\_diff                     & 0.0   & +0.953        & dif\_too               & 1.0   & +1.155        & Bias term & 1.0   & +0.872        \\
	Bias term & 1.0   & +0.419        & Bias term & 1.0   & +0.594        & has\_int                     & 0.0   & +0.450        \\
	dif\_but               & 0.0   & +0.114        & dif\_consider          & 1.0   & +0.491        & dif\_may               & 1.0   & +0.393        \\
	dif\_because           & 0.0   & +0.058        & dif\_however           & 1.0   & +0.186        & dif\_but               & 1.0   & +0.306        \\
	dif\_how               & 0.0   & +0.033        & dif\_how               & 1.0   & +0.092        & dif\_hope              & 1.0   & +0.215        \\
	dif\_yet               & 0.0   & +0.033        & dif\_hope              & 0.0   & +0.027        & dif\_while             & 0.0   & +0.060        \\
	int\_unity             & 0.0   & +0.029        & dif\_perhaps           & 0.0   & +0.019        & dif\_rather            & 0.0   & +0.058        \\
	dif\_depend            & 0.0   & +0.024        & dif\_almost            & 0.0   & +0.018        & dif\_too               & 0.0   & +0.033        \\
	dif\_hope              & 0.0   & +0.022        & dif\_sometimes         & 0.0   & +0.012        & dif\_seem              & 0.0   & +0.030        \\
	dif\_rather            & 0.0   & +0.022        & dif\_although          & 0.0   & +0.011        & dif\_differ            & 0.0   & +0.026        \\ \midrule
	has\_int                     & 0.0   & -0.009        & dif\_while             & 0.0   & -0.032        & int\_remain            & 0.0   & -0.024        \\
	dif\_close\_to         & 0.0   & -0.009        & dif\_rather            & 0.0   & -0.033        & dif\_separate          & 0.0   & -0.026        \\
	dif\_seem              & 0.0   & -0.010        & dif\_different         & 0.0   & -0.050        & int\_weigh             & 0.0   & -0.028        \\
	dif\_consider          & 0.0   & -0.011        & dif\_often             & 0.0   & -0.050        & dif\_possible          & 0.0   & -0.029        \\
	int\_account           & 0.0   & -0.012        & dif\_each              & 0.0   & -0.050        & int\_unity             & 0.0   & -0.031        \\
	dif\_secret            & 0.0   & -0.013        & dif\_either            & 0.0   & -0.069        & dif\_however           & 0.0   & -0.036        \\
	dif\_differ            & 0.0   & -0.018        & dif\_about             & 0.0   & -0.078        & dif\_often             & 0.0   & -0.037        \\
	dif\_usually           & 0.0   & -0.019        & dif\_both              & 0.0   & -0.079        & dif\_about             & 0.0   & -0.068        \\
	int\_remain            & 0.0   & -0.024        & dif\_because           & 0.0   & -0.284        & dif\_though            & 0.0   & -0.073        \\
	dif\_may               & 0.0   & -0.036        & dif\_but               & 1.0   & -0.575        & dif\_because           & 0.0   & -0.102        \\ \bottomrule
	\end{tabular}%
}
\caption{Top ten and bottom ten features used in successful IC classifications using vocabulary features. Differentiation and integration terms are prefixed with \emph{dif} and \emph{int}, while has\_dif and has\_int are the binary features for whether any differentiation/integration terms are present at all. The bias term is the averaged sum of the value associated with each root node in the ensemble.}
\label{classify_vocab}
\end{table*}

\subsection{Feature analysis}

We now examine the role played by semantic and syntactic features. For the Vocabulary and V+POStags feature sets, we look at three aspects. First, which features are \emph{never} used by any decision trees in the ensemble. Second, the importance attached to features which \emph{are} used as measured by averaging the information gain of the feature across all decision trees in the ensemble. Last, in instances of correct classifications by a model, examples of which features are most discriminative for a particular IC band.

\vspace{4pt}\noindent\textbf{Vocabulary.} Of the 313 features, only 89 are used. Unused features are either extremely common words or not present at all. The majority of top features are differentiation keywords. The Vocabulary model correctly classified test items in three IC bands only (Table~\ref{classify_vocab}). For band 1, the most important feature is that denoting the absence of any differentiation-related words. For band 2, this same feature also plays a role though does not appear in the top 10. Here, the presence of the differentiation terms themselves are of more importance. And for band 3, the binary feature for integration-related vocabulary, set to 0, is most important alongside the differentiation keywords. The failure of the Vocabulary model to correctly classify the band 4 text is easily explained: it contains very few of the expected vocabulary items.

\vspace{4pt}\noindent\textbf{V+POStags.} A total of 98 features are used. Of the 45 POS features, 10 are never used because too frequent or too rare. Only 63 vocabulary items are used, compared to the 89 used in the vocabulary-only model. Overall, the most important syntactic features are adjectives and adverbs. This is not surprising, given that the majority of differentiation/integration keywords are in this category. Predeterminers, which are found in phrases such as ``all this'', ``what a'' and ``many times the'', are also important. Even though semantic features result in the greatest information gain during training, syntactic features turn out to be the most useful in the evaluation. Table~\ref{classify_vpos} shows the features which most contributed to a correct classification. The band 1 text lacks many syntactic features (shown as having a value of 0.000) and even though there is some evidence of differentiation, these features are negatively weighted. This may be attributed to the fact that the presence of keywords is not necessarily evidence of IC. For IC bands 2 and 3, the distinct lack of integration features is important, along with the presence of some syntactic features. The band 4 text, by comparison, has fewer zero-value features and may be considered a more complex text in terms of syntax, which here also belies its higher level of IC.

\begin{table*}[t]
\centering
\resizebox{1.99\columnwidth}{!}{%
	\begin{tabular}{@{}rccrccrccrcc@{}}
	\toprule
	\multicolumn{3}{c}{IC = 1}                            & \multicolumn{3}{c}{IC = 2}                            & \multicolumn{3}{c}{IC = 3}                            & \multicolumn{3}{c}{IC = 4}                            \\ \midrule
	Feature                       & Value & Contribution & Feature            & Value & Contribution  & Feature                       & Value & Contribution & Feature                       & Value & Contribution \\ \midrule
	Bias term               & 1.000 & +0.894        & Bias term 			   & 1.000 & +0.786        & Bias term              & 1.000 & +1.150        & Noun, singular                 & 0.194 & +0.506        \\
	Verb, present                & 0.000 & +0.672        & dif\_too         & 1.000 & +0.743        & Verb, base form                 & 0.130 & +0.308        & Adj.                 & 0.081 & +0.441        \\
	Coord. conj.                 & 0.022 & +0.642        & Verb, past                & 0.000 & +0.208        & Noun, plural                & 0.056 & +0.235        & Subord. conj.                 & 0.145 & +0.394        \\
	Particle ``to''                 & 0.000 & +0.513        & Adj.                 & 0.051 & +0.207        & Noun, singular                 & 0.056 & +0.230        & Verb, base form                 & 0.016 & +0.340        \\
	Preposition                & 0.000 & +0.497        & Noun, singular                 & 0.120 & +0.143        & Subord. conj.                 & 0.074 & +0.132        & Verb, past part.                & 0.016 & +0.336        \\
	Verb, past                & 0.067 & +0.338        & Verb, base form                 & 0.043 & +0.129        & Verb, past part.                & 0.000 & +0.129        & Comp. adj.                & 0.016 & +0.186        \\
	Proper noun                & 0.222 & +0.317        & Preposition                & 0.060 & +0.121        & Coord. conj.                  & 0.037 & +0.109        & Verb, n3ps pres.                & 0.016 & +0.163        \\
	Subord. conj.                 & 0.111 & +0.259        & Possessive                & 0.000 & +0.052        & dif\_but         & 1.000 & +0.081        & Determiner                 & 0.081 & +0.149        \\
	Determiner                 & 0.111 & +0.161        & Modal verb                 & 0.034 & +0.030        & has\_int               & 0.000 & +0.078        & dif\_but         & 1.000 & +0.136        \\
	Comp. adj.                & 0.000 & +0.136        & has\_int               & 0.000 & +0.027        & dif\_too         & 0.000 & +0.066        & Verb, present                & 0.000 & +0.133        \\ \midrule
	Modal verb                 & 0.000 & -0.020        & Noun, plural                & 0.026 & -0.113        & Verb, n3ps pres.                & 0.056 & -0.098        & dif\_yet         & 0.000 & -0.034        \\
	Adv., superlative                & 0.000 & -0.021        & Determiner                 & 0.094 & -0.129        & Proper noun                & 0.000 & -0.120        & Particle ``to''                 & 0.016 & -0.035        \\
	Particle                 & 0.000 & -0.060        & Adverb                 & 0.060 & -0.133        & Verb, past                & 0.000 & -0.122        & Existential ``there''                 & 0.000 & -0.047        \\
	has\_diff          & 1.000 & -0.251        & Verb, past part.                & 0.017 & -0.133        & Adverb                 & 0.093 & -0.142        & dif\_how         & 0.000 & -0.063        \\
	Adverb                 & 0.022 & -0.255        & Verb, present                & 0.026 & -0.137        & Preposition\$              & 0.019 & -0.152        & Verb, gerund                & 0.048 & -0.077        \\
	Verb, gerund                & 0.022 & -0.281        & Verb, n3ps pres.                & 0.017 & -0.205        & Modal verb                 & 0.074 & -0.178        & Coord. conj.                  & 0.065 & -0.134        \\
	Verb, past part.                & 0.000 & -0.342        & Proper noun                & 0.017 & -0.296        & Particle ``to''                 & 0.056 & -0.262        & Wh-adverb                & 0.000 & -0.196        \\
	Verb, n3ps pres.                & 0.000 & -0.501        & Possessive pronoun              & 0.043 & -0.341        & Adj.                 & 0.111 & -0.268        & Modal verb                 & 0.000 & -0.258        \\
	Adj.                 & 0.111 & -0.716        & Subord. conj.                 & 0.162 & -0.341        & Determiner                 & 0.056 & -0.282        & Noun, plural                & 0.113 & -0.524        \\
	Wh-determiner                & 0.022 & -0.861        & Adj., superlative                & 0.017 & -1.046        & Verb, present                & 0.000 & -0.315        & Bias term & 1.000 & -0.818        \\ \bottomrule
	\end{tabular}%
}
\caption{Top ten and bottom ten features used in successful IC classifications using V+POStags features. The bias term is the averaged sum of the value associated with each root node in the ensemble.}
\label{classify_vpos}
\end{table*}

\subsection{Sensitivity analysis}

To adhere to the original theoretical framework as much as possible, our operationalization of IC follows a 7-class scale. In addition to that, we check the robustness of our results to coarser IC aggregations, also motivated by previous work in which fewer and wider bands of IC were used \cite{ambili2014automated}. We first tried a 4-way classification: no IC (class 1 only), low IC (classes 2 and 3 joint), medium IC (4 and 5), high IC (6 and 7). We then collapsed the low and medium classes to try a ternary classification. We always kept class 1 separate (and did not merge it with class 2, for example) because it has a very distinctive meaning: it is the only class that exhibits no sign of differentiation or integration.

\begin{figure}[t!]
	\centering
		\includegraphics[width=0.45\columnwidth]{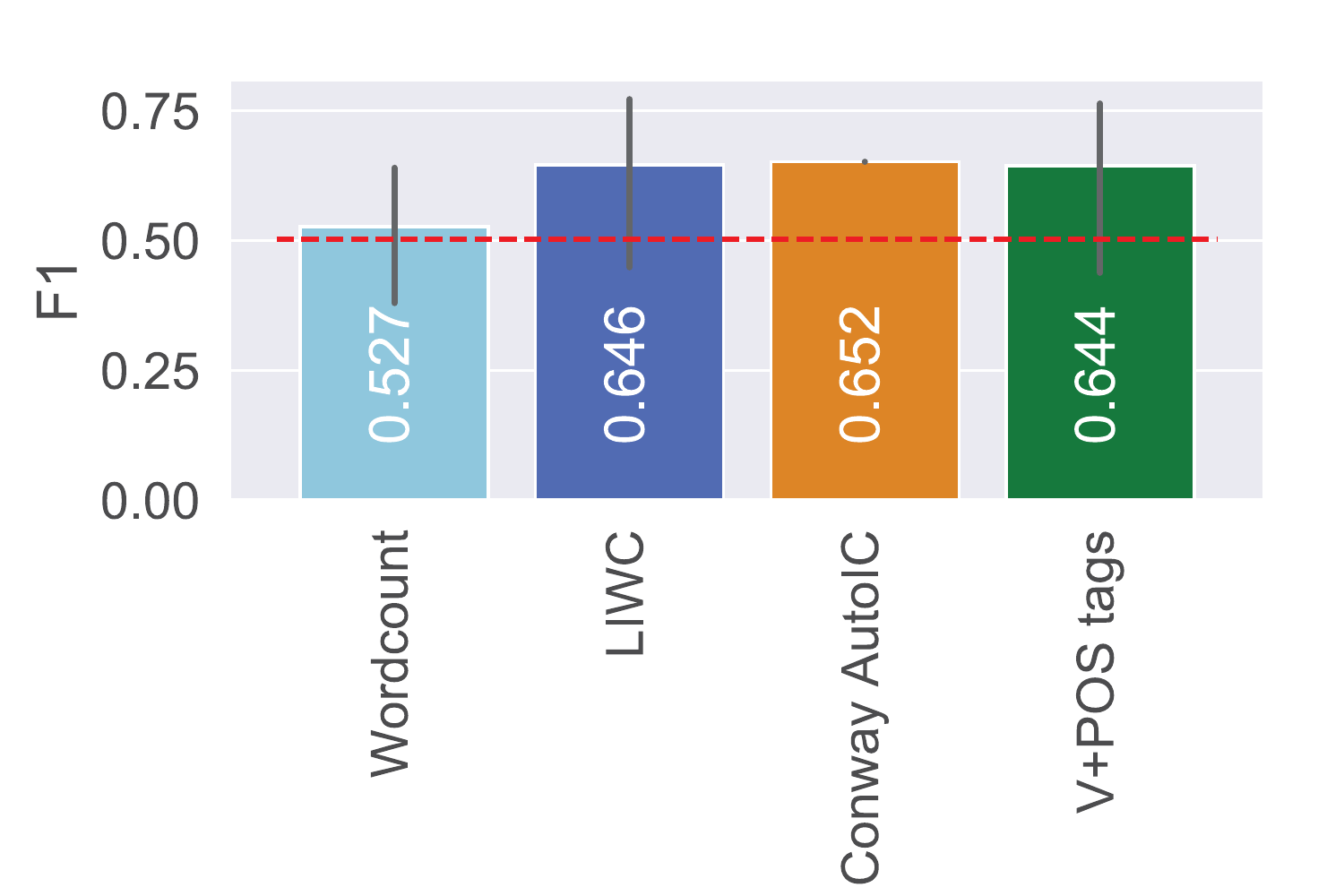}
		\includegraphics[width=0.45\columnwidth]{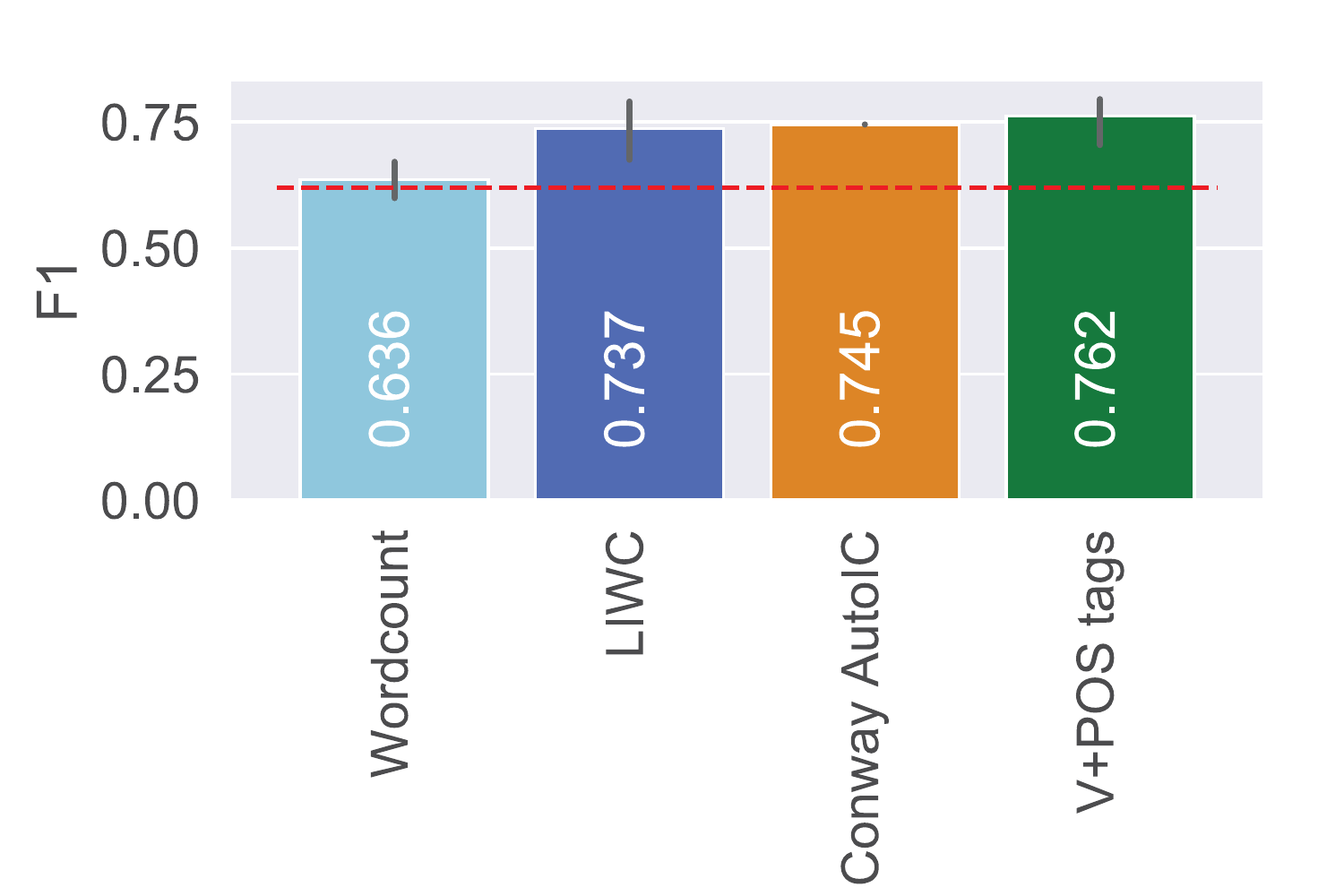}\\
		\includegraphics[width=0.45\columnwidth]{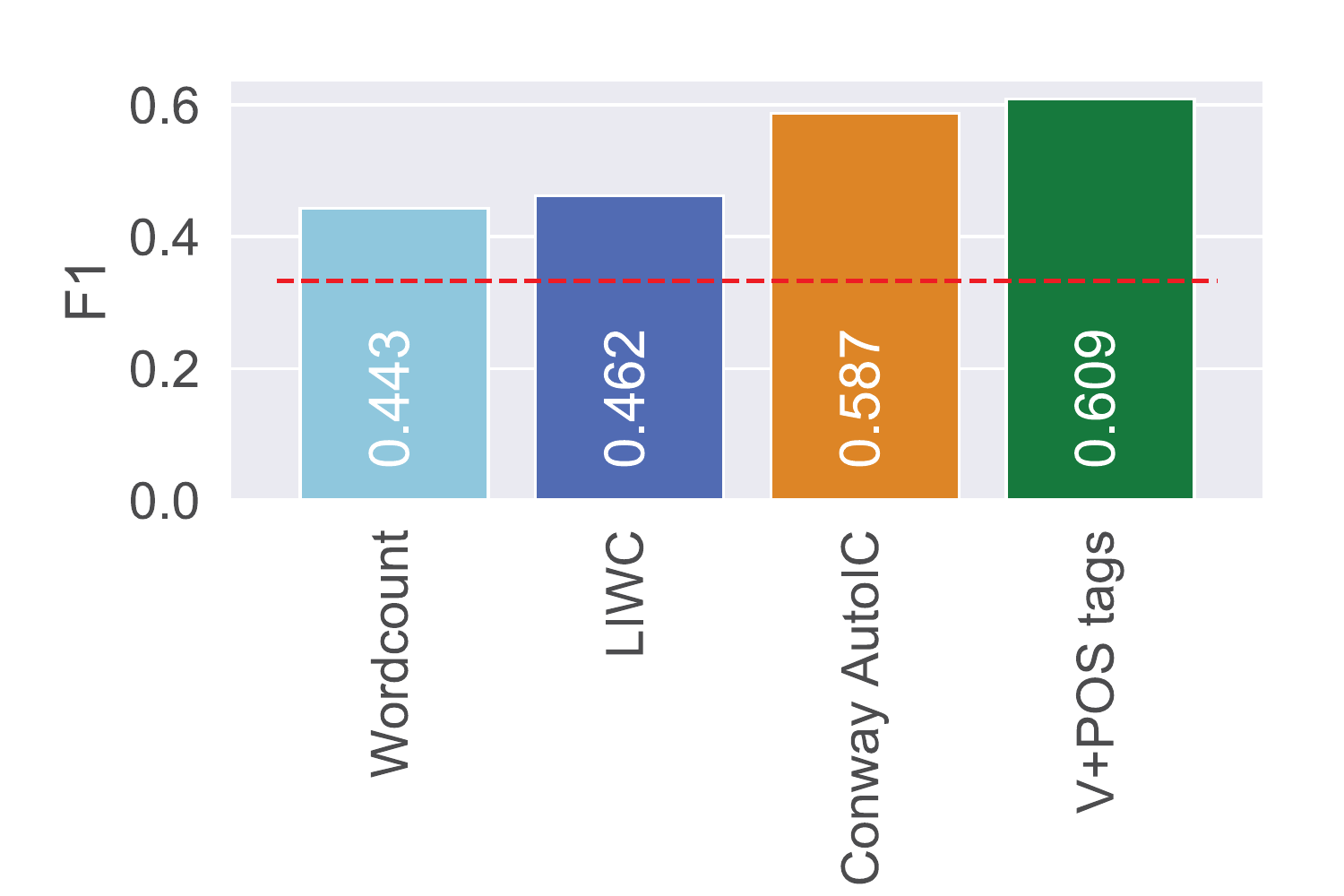}
		\includegraphics[width=0.45\columnwidth]{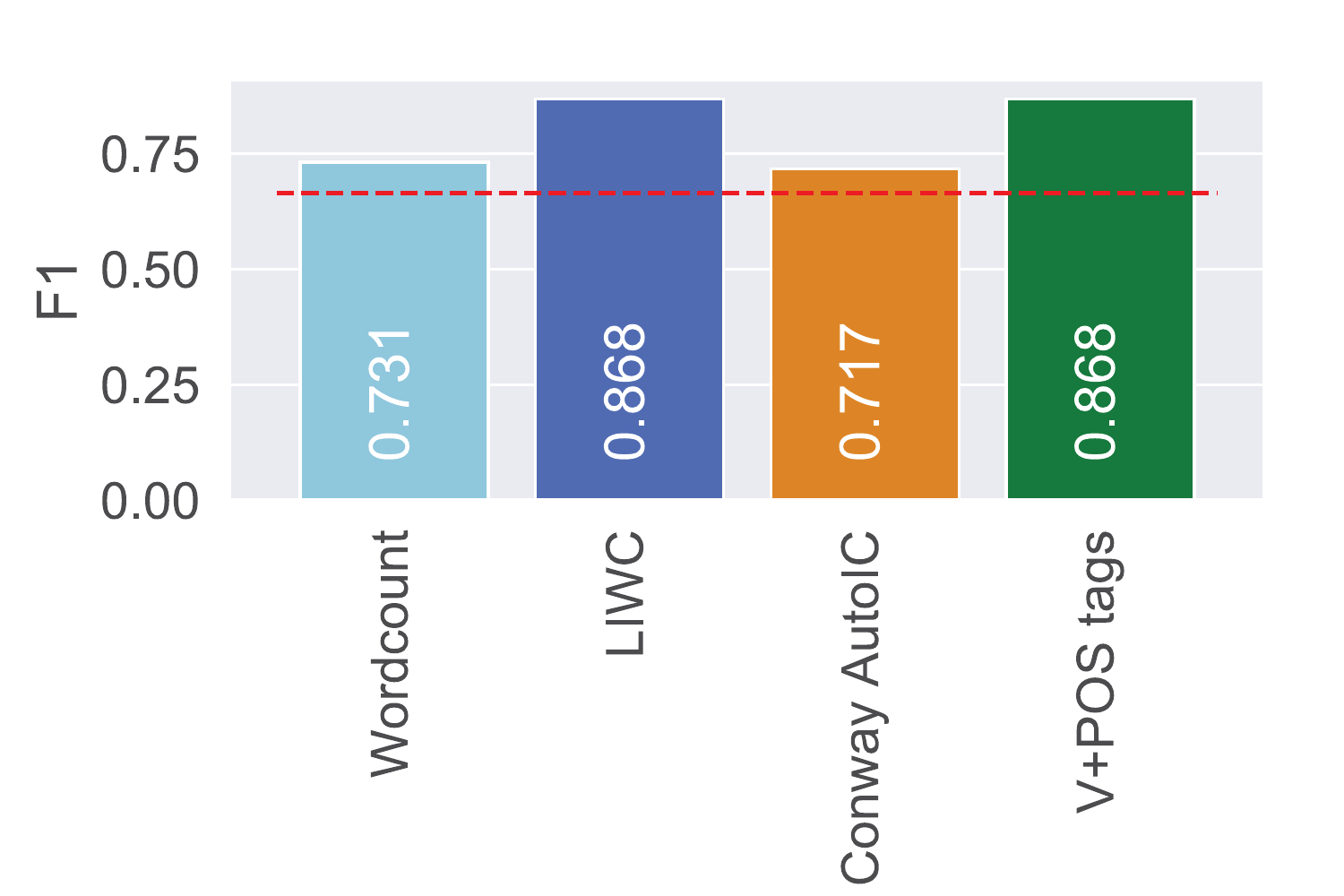}
	\caption{Classification results by aggregating the 7 IC bands into fewer classes. Left: 4-class classification (1, 2+3, 4+5, 6+7). Right: 3-class classification (1, 2+3+4+5, 6+7). Top: results of cross-validation. Bottom: results of classification on the heldout dataset.}
	\label{fig:class_aggregation}
\end{figure}

Results are presented in Figure~\ref{fig:class_aggregation}. For brevity, we report the results for the main baselines and for the best performing model. The pattern is consistent with that of the 7-band classification. Word count is the weakest approach; LIWC, Conway's IC, and V+POS Tags perform comparably in cross validation; V+POS Tags is always the best approach in the classification on the heldout dataset.

\begin{figure}[t!]
	\centering
	  \includegraphics[width=0.49\columnwidth]{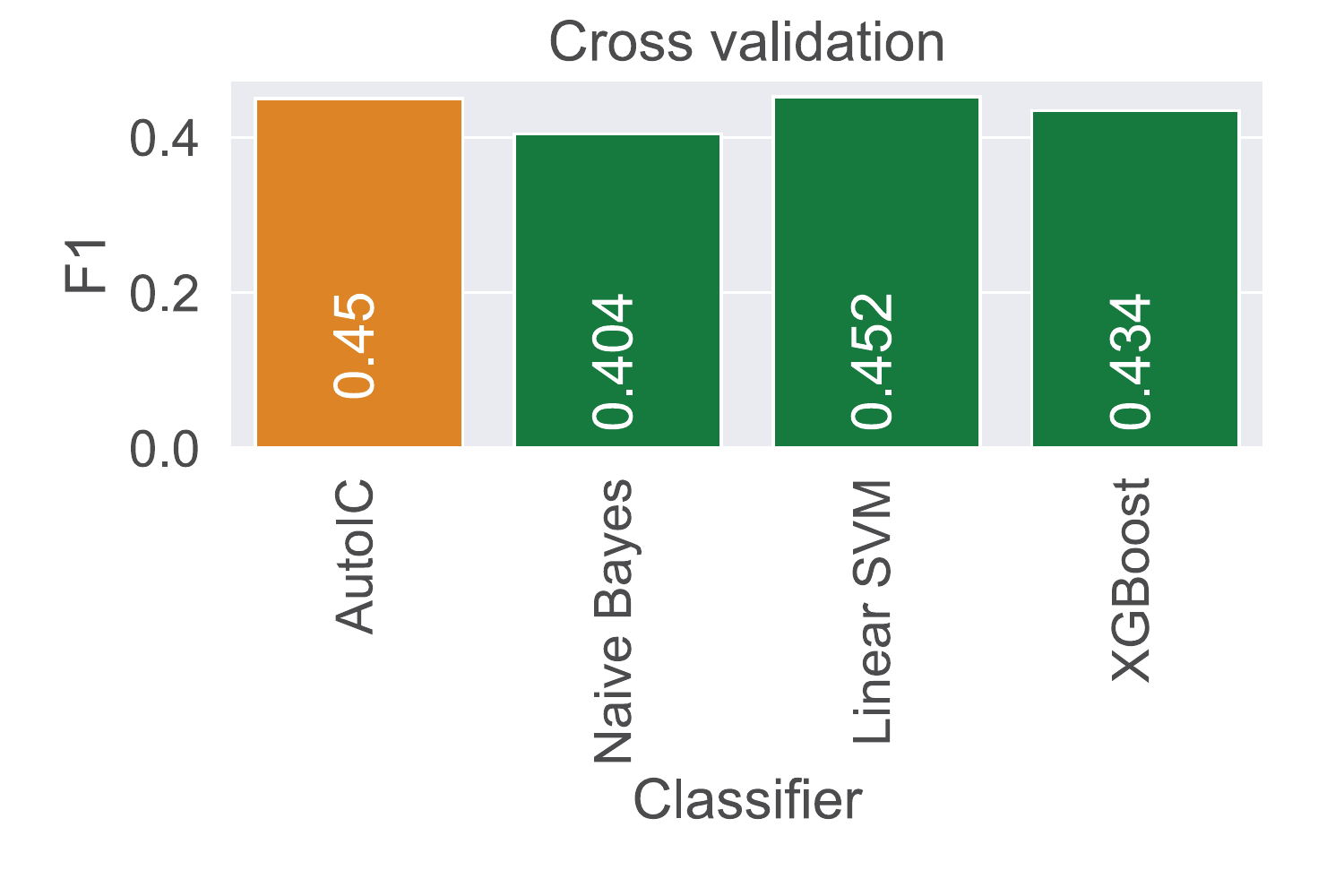}
		\includegraphics[width=0.49\columnwidth]{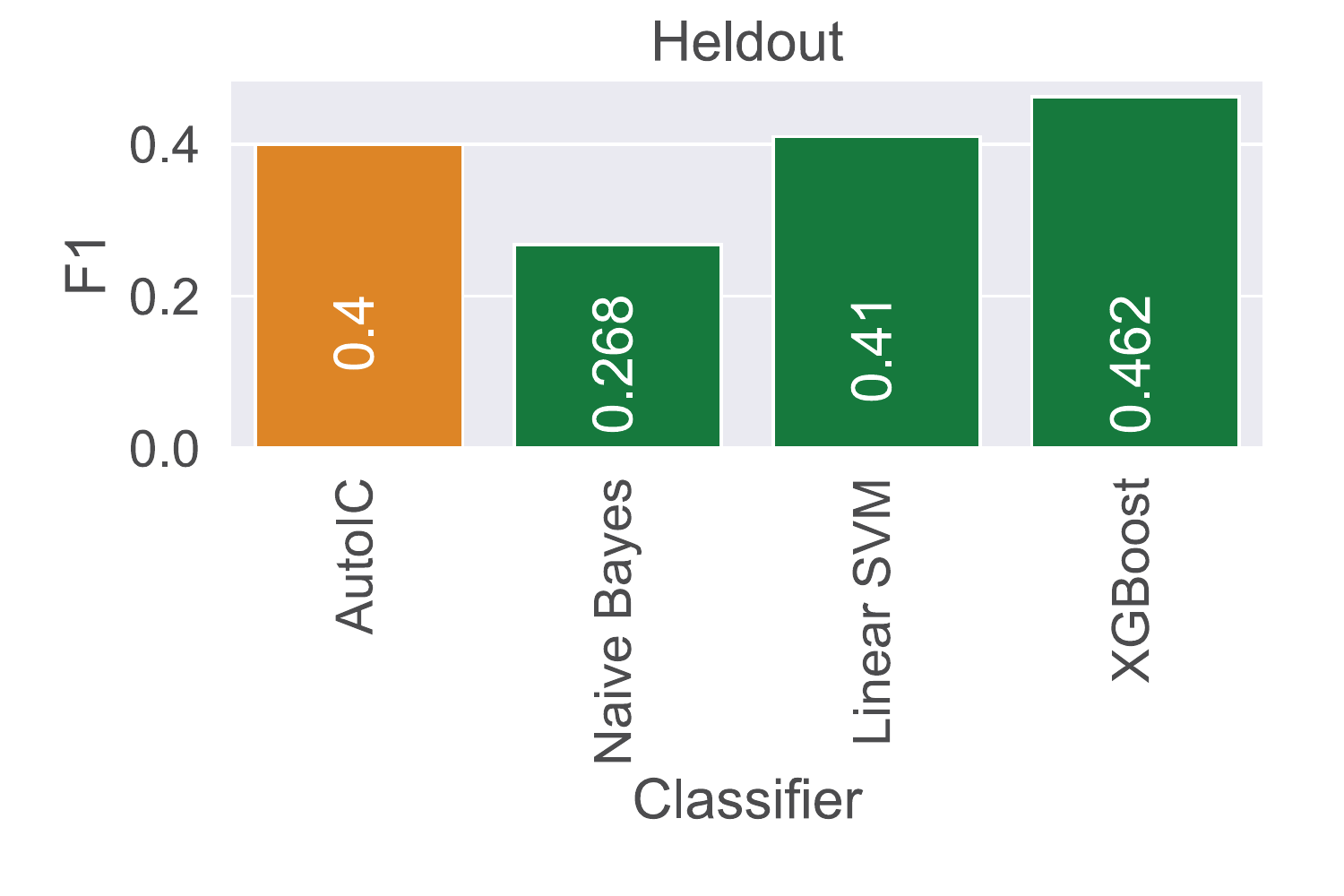}
	\caption{Comparison between classifiers in cross validation (left) and heldout classification (right) using the V+POS Tags features. AutoIC is reported for comparison.}
	\label{fig:classifiers_comparison}
\end{figure}

In addition, we also experiment with other classification models to check the robustness of our approach. We compare xgboost with Naive Bayes as a simple baseline and with linear SVM, which achieves better generalization than decision trees. Results are reported in Figure~\ref{fig:classifiers_comparison}. As expected, Naive Bayes is the weakest approach. Linear SVM is second to xgboost and keeps a rather decent performance, being able to slightly beat AutoIC both in the cross validation and heldout setups. 

\subsection{Classification summary}

A purely syntax-based approach to IC scoring, which is well-motivated given the theory behind IC, performs reasonably well on held-out evaluation data. However, this is not sufficient to surpass the performance of the current state of the art as represented by the semantic-based approach of Conway's AutoIC tool. This is only possible by combining syntax and semantics together in one model. This allows the model to look beyond the surface forms of language, the \emph{what} of what is being said, and leverage the syntactic properties of the text to access the \emph{how} of what is being said. Limited training examples and stylistic differences within the training data that is available accounts for why syntax alone is insufficient and must be bootstrapped, to some degree, by a lexicon of differentiation and integration. The best-performing model represents an almost 25\% improvement over the current state of the art. As a contribution to the community, we make our model public and open-source (social-dynamics.net/ic).

\section{Measuring IC in social media}\label{sec:social}

After testing our model on annotated data, we apply it on a larger set of posts and comments from Reddit with the goal of building initial evidence about its external validity. Based on previous literature, we set an hypothesis about the level of IC that we expect to find under certain conditions. This is the first time an analysis of Integrative Complexity is conducted at this scale. Although Conway's AutoIC tool has been used in research~\cite{oscars,piggy}, it has not been tested in this manner and on large-scale data.

\subsection{Data}

Reddit is a social media site focused on news aggregation and user discussion. It is organized into themed \emph{subreddits} where users submit posts for others to both comment and vote on. We focus on the three subreddits in Table~\ref{redditdesc}, which allow only text-based posts and are focused on particular forms of discussion. Using the Reddit API to collect all posts and comments made between January 2018 and August 2018, we gathered data from /r/depression, a support-based subreddit focused on mental health, as well as two subreddits where the focus is on high quality responses to specific questions. The lower number of posts/comment in /r/AskHistory and /r/AskScience is due to both stricter controls on quality of submissions and a much narrower focus.

\begin{table}[t]
\centering
\resizebox{0.95\columnwidth}{!}{%
\begin{tabular}{rccc}
\hline
Subreddit & Subscribers & Posts & Comments \\ \hline
/r/depression & 380k & 8k & 212k \\
/r/AskHistorians & 790k & 1.6k & 42k \\
/r/AskScience & 15.9m & 1.2k & 143k \\ \hline
\end{tabular}%
}
\caption{Data collected from Reddit.}
\label{redditdesc}
\end{table}

\subsection{Hypothesis}

Findings of \citeauthor{suedfeld1993changes} \citeyear{suedfeld1993changes} show that IC for personal writings made during periods of severe personal distress (e.g., following the death of a loved one or a betrayal) was higher than those written under normal conditions, before the negative event. Positive events (e.g., marriage, career success) had no effect. Based on these findings, we hypotesize that texts from the /r/depression subreddit, where users write about their experience of depression, often triggered by difficult personal circumstances, grief, and other traumas~\cite{de2014mental}, exhibit higher IC than what is measured in other discussions about non-dysphoric experiences. We therefore compare texts from /r/depression with other two communities, /r/AskScience and /r/AskHistorians, which are focused on knowledge exchange rather than sharing negative experiences and providing social support. As the latter two subreddits include threads rich of content and debates about complex and possibly controversial themes (e.g., how the universe was originated and will end), we expect them to contain a non-negligible number of posts with some level of integrative complexity. However, because the theory predicts that being in a state of psychological distress adds an additional layer of complexity to a person's reasoning, we predict that texts from /r/depression will exhibit markedly higher IC scores.

\subsection{Results}

\begin{figure}[t!]
	\centering
	  \includegraphics[width=0.49\columnwidth]{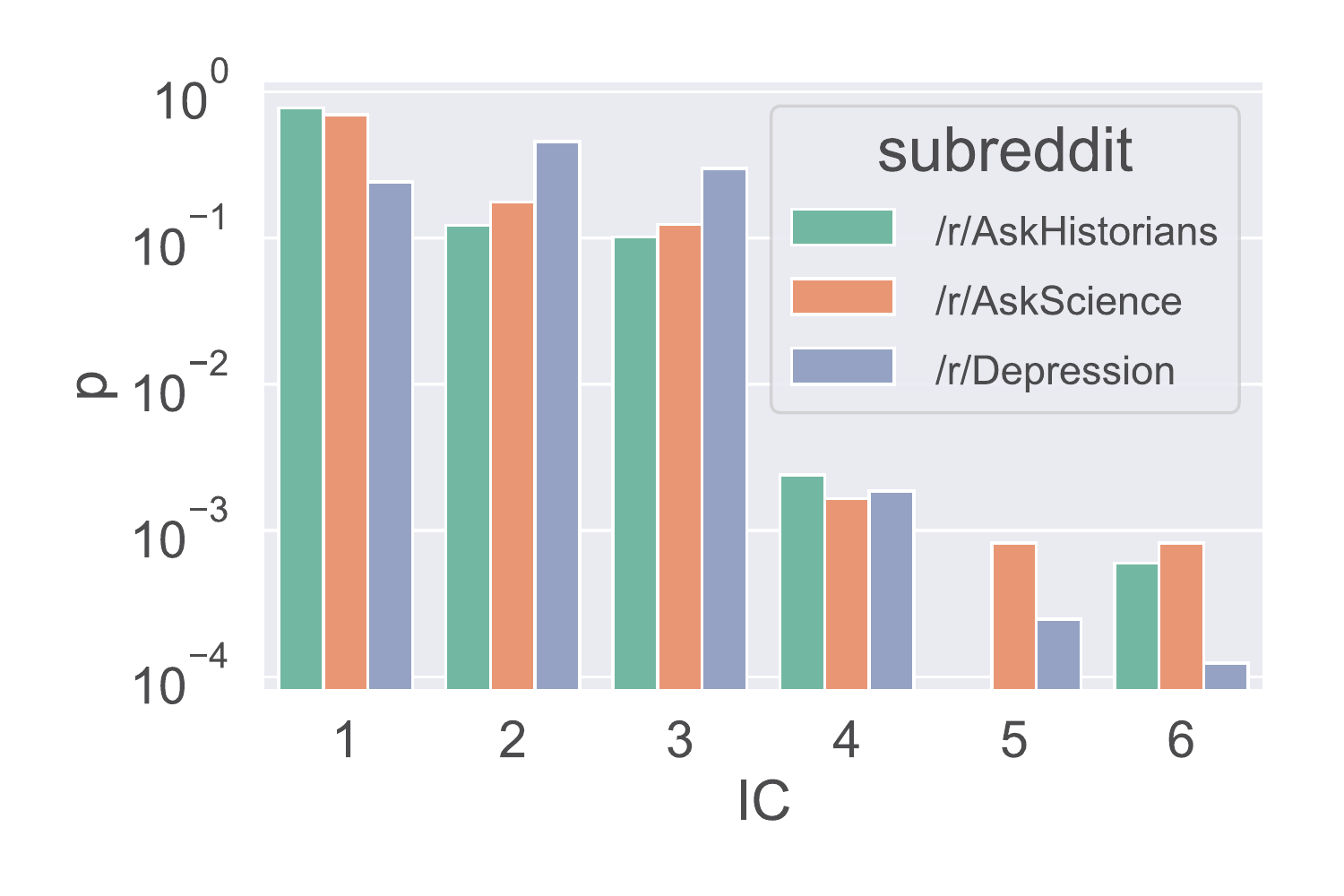}
		\includegraphics[width=0.49\columnwidth]{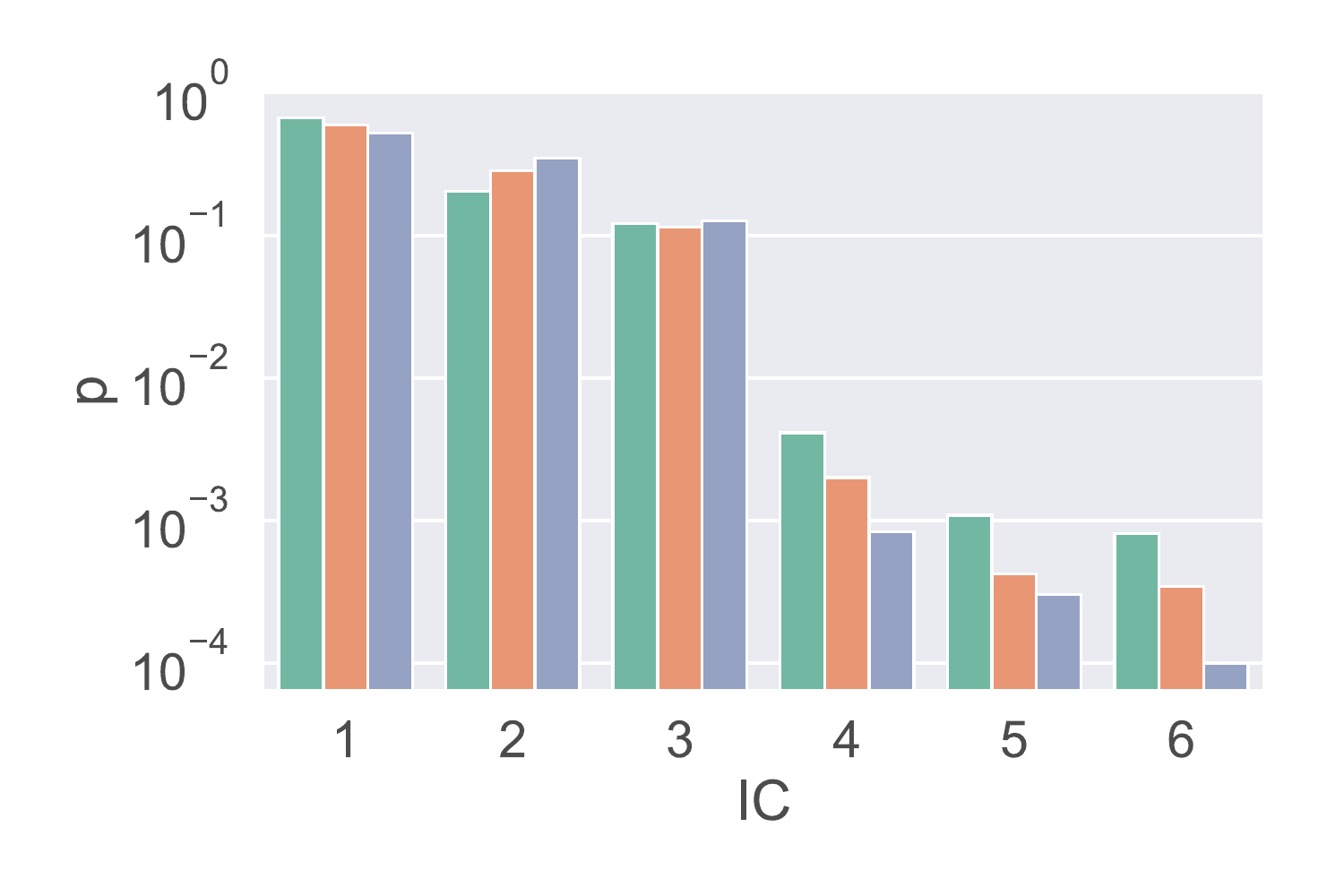}
	\caption{Probability distribution of IC values in the posts (left) and comments (right) of the three subreddits considered.}
	\label{fig:distr_ic_posts_comments}
\end{figure}

We apply the best-performing classifier (V+POStags features), trained on all available training data, to the post and comment text of the subreddits. The resulting probability distributions over IC scores is shown in Figure~\ref{fig:distr_ic_posts_comments}. Even if posts with very high scores are globally rare, we found a variety of scores in all the subreddits. Not surprisingly, the majority of Reddit posts analyzed belong to class 1---no sign of IC at all. The distribution in /r/depression is more right-skewed than those from the other two subreddits, which provides a first indication of higher IC levels in depression-related discussions. The average IC scores in posts and comments, per subreddit, are summarized in Table~\ref{chi_results}. Examples of high and low IC texts for the different online communities are shown in Table \ref{example_texts_reddit}.

\begin{table}[tp]
\centering
\resizebox{0.95\columnwidth}{!}{%
\begin{tabular}{rcc}
\toprule
\multicolumn{1}{c}{Subreddit}  & Posts            & Comments                            \\
\midrule
/r/depression        & $\mu$ 2.06 ($\sigma$ 0.74)      & $\mu$1.61 ($\sigma$ 0.71)      \\
/r/AskHistorians     & $\mu$ 1.34 ($\sigma$ 0.67)      & $\mu$1.47 ($\sigma$ 0.74)      \\
/r/AskScience        & $\mu$ 1.44 ($\sigma$ 0.73)      & $\mu$1.52 ($\sigma$ 0.71)      \\
\bottomrule
\end{tabular}%
}
\caption{Mean IC scores for posts and comments, per subreddit.}
\label{chi_results}
\end{table}

\begin{table}[tp]
\centering
\resizebox{0.95\columnwidth}{!}{
\begin{tabularx}{1.4\columnwidth}{@{}llX@{}}
\toprule
Source & IC & Text \\
\midrule
Depression & H & I am so lucky to have friends who understand me. But I regret telling my recent ex about my depression. He used to understand, but I mean I get it. He was mentally unstable too.\\
Depression & L & A lot of people are lurkers. Doesn't mean they don't care, they just don't know what to say and sometimes that's better than saying something bad. Just write whatever you want, like let it all out, maybe someone out there feels the same way and they just didn't want to write it too. \\
\midrule
AskHistory & H & The Lewis and Clark parallel is spot on. Not everyone in 1830s America was wilderness-capable, but the percentage would be much higher than it is today. Hunting and foraging was common in the frontier, which in this era started around Wisconsin, Illinois, and west. It's also not too hard for a single person with a rifle and a fair amount of survival skills to scrabble through in the wild. Lansford Hastings was a politician in the Republic of California - he knew just  enough trailblazing to sell a plausible trail, but not to ensure its safety. \\ 
AskHistory & L & So I'm looking to teach myself an ancient language, and would like to translate text that has yet to be translated yet. Are there any projects out there that need help? \\
\midrule
AskScience & H & Nonsense.  Both your esophagus and trachea are valve protected (and the pressure would be preferentially released through your nose and lips but even if it wasn't) there are sphincters around your stomach and there is your anal sphincter and all of these are well able to retain an atmosphere of pressure. If anything your difficult in this scenario would be exhaling without atmospheric pressure to help your diaphragm and supporting muscles cause your rib cage to contract. \\
AskScience & L & What happens exactly with the stability of therapeutic proteins when kept at room temperature? \\
\bottomrule
\end{tabularx}}
\caption{Examples of Low ($<$3) and High ($\geq$3) IC texts from specific subreddits.}
\label{example_texts_reddit}
\end{table}

\begin{figure*}[t!]
	\centering
		\includegraphics[width=1.70\columnwidth]{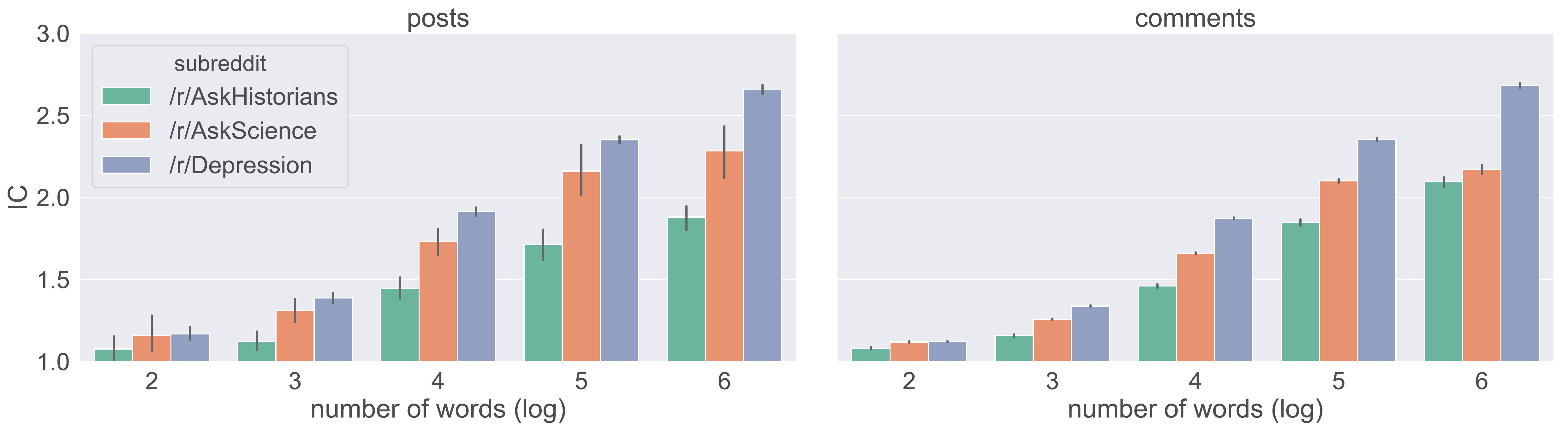}
	\caption{Average IC of posts (left) and comments (right) in the three subreddits considered, binned by text length (log of the number of words, rounded to the next integer). Depression-related posts and comments have higher IC compared to texts of comparable length from the other two subreddits. Bars show the 95\% confidence intervals of the mean.}
	\label{fig:reddit_posts_ic}
\end{figure*}

Our classifier does not take into account text length to estimate IC classes and, as a result, short texts can be scored as highly complex. In Table~\ref{example_texts_reddit}, for example, we report two comments from /r/depression where the shorter one has much higher complexity than the longer one. However, it is known that text length might correlate with IC~\cite{baker1990coding}. It is harder to compress a lot of information about differentiation and integration into a short piece of text and, as a result, longer texts will be associated with higher IC, on average. To make sure that the difference in average IC across subreddits is not due only to differences in text length, we contrast the level of IC of posts and comments across subreddits for texts of comparable length (Figure~\ref{fig:reddit_posts_ic}). Consistent with what theory would suggest, we observe that depression-related posts have higher IC than history- or science-related posts. A similar trend emerges for comments. The gap between them is not prominent for very short posts (those roughly under 10 words, represented in the first bin in Figure~\ref{fig:reddit_posts_ic}). As the text length increases, the overall levels of IC rise but sensibly more so for depression-related posts. Although text length and IC correlate in this specific Reddit sample, large differences in IC levels for texts with similar length do exist.

\begin{figure*}[t!]
\centering
	\includegraphics[width=1.80\columnwidth]{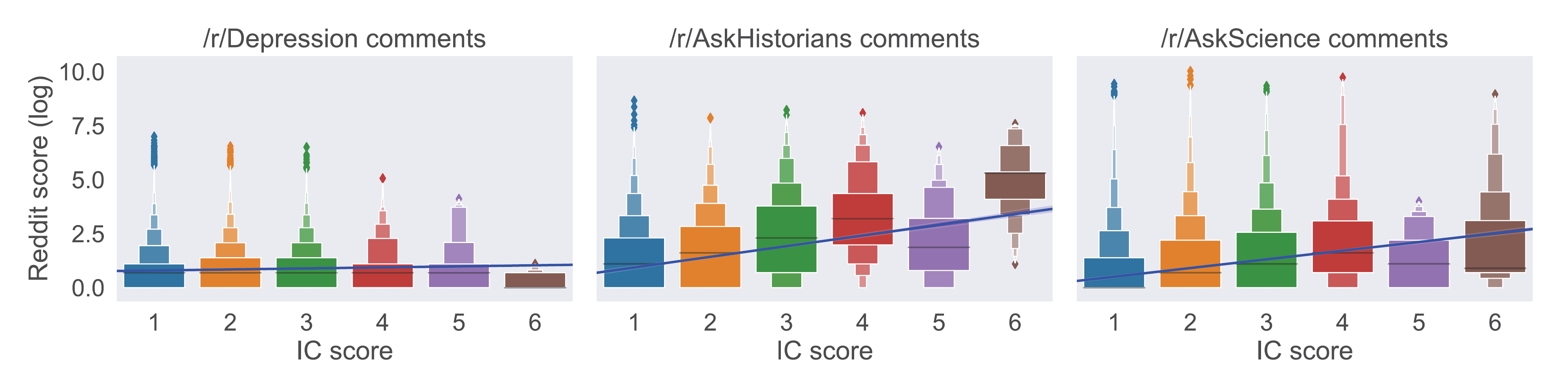}
	\caption{Value plots showing the distribution of Reddit comment scores for each IC band, for each subreddit. The black lines shows the mean, with each box representing a percentile. The blue regression line shows the linear relationship between IC and Reddit score.}
	\label{reddit_results}
\end{figure*}

Last, in addition to looking at the differences in IC across subreddits, we also look at how users reactions are associated to these differences as a first example of exploratory analysis that could be enabled by the application of our tool at scale.

In Reddit, the community votes on content, with popular content having a higher score than bad. Figure~\ref{reddit_results} shows the distribution of Reddit scores per IC band. Because of the very long tail of these scores, we use a value plot \cite{hofmann2017value}. This non-parametric visualisation displays more percentiles than a standard boxplot, with each percentile represented as a box whose height denotes the range of values within that percentile. To emphasize the trend we display a linear regression line. In /r/AskHistorians and /r/AskScience, better answers tend to be more integratively complex. These are competitive subreddits where users spend time writing informative comments which meet the rules of the community and answer the question at hand and complex answers are likely to be rewarded. On the contrary, /r/depression is a support subreddit where there is no explicit motivation to reward complex comments:  the main purpose of comments in this subreddit is that of social support rather than of discussion enrichment with new perspectives~\cite{de2014mental}.

\section{Discussion and conclusion}\label{sec:conclusion}

The language of dialogue is complex: people who are able to recognise the differences between conflicting points of view and to identify paths for their reconciliation use not only the right words but they combine them wisely. We capture that by designing a system for automatic scoring of Integrative Complexity that blends semantic and syntactic features. Its accuracy is sufficient to replicate findings from prior work, showing that people experiencing personal crisis exhibit higher levels of IC than those who are not. This validated advance will hopefully encourage the application of this measure of cognitive processes to a wider range of situations. Next, we discuss some of the limitations of our work and some of the opportunities it opens up.

\subsection{Implications}

This work results in theoretical and practical implications. From the theoretical standpoint, this work reinforces the evidence that IC can be operationalized and that it can be done most effectively when language syntax is brought into the equation. From a methodological perspective, by casting the IC scoring task as a classification problem and generating labels rather than real-valued output, as in \citeauthor{conway2014automated} \shortcite{conway2014automated}, our approach more closely follows the methodology laid out in the IC coding manual \cite{baker1990coding}. 

By opening our method to the research community, we start paving the way towards important practical applications in social media analytics. In the last decades, thanks to the diffusion of tools to calculate psychological attributes easily and fast, the application of psychometrics have expanded its scope from small laboratory experiments to large-scale studies with online data. The shortened Big-5 personality test~\cite{gosling2003very} and the LIWC dictionary~\cite{liwc} are just some examples. Our method aims at complementing these type of approaches by providing the first open tool to calculate IC fast, at scale, and on any type of text. Our tool does not require subjects to take a task-specific test and does not rely completely on wordlists and can therefore detect cognitive processes quite flexibly. In our Reddit study, we were able to code around 400k texts in hours---a task that would take human coders an incredible amount of time.

\begin{table*}[t!]
	\centering
\resizebox{1.95\columnwidth}{!}{%
		\begin{tabular}{cp{16cm}}
			IC & Text  \\ \midrule
 1 & I have strong negative feelings towards this topic in a society where we are supposed to be somewhat evolved socially and economically. Killing a man for his crimes does not seem to be the most evolved or consistent thing. To do an eye for an eye has existed for ages and you think the United States would be over that by now. \\
2 & My opinion is negative because the death penalty attempts to right a wrong with a similar wrong action. A person should be punished by other means which are more humanistic and thoughtful. \\
3 & I think that being assertive is important but at the same time being too assertive can be a bad thing. Being assertive can get things done in a much more effective manner. I am personally an assertive person and think that it can lead to problems with others. However, just because a person is not an assertive person doesn't mean that they should possess that quality. Every person is different and has different characteristics.  \\
4 & I am a loving, caring, humorous individual who can be vicious, vindictive, and thoughtless. at times i'm impatient, insightful, and demanding. I recognize what can be and what is and usually am able to suggest how to bridge the gap. I feel that I have learned these qualities from my family and experience with the world. \\
5 & My family existed during the great depression of the 1930's and like most people we lived on very little money. No welfare or pensions then but lots of love and sharing with others. That, with my nursing experiences later and friendships contributed to my present firm belief that money and possession of things do not necessarily result in happiness. \\
6 & I grew up in Canada in a medium sized white middle class family. As such part of what makes me who I am are the values and beliefs I was taught, and have subsequently questioned, of a western patriarchical society. I very much understand myself to be part of a larger scheme of things partly a product of my culture and of my upbringing. However, I also see myself as an individual a product of my own unique experiences and views. So I guess what makes me ``me'' is the culture I come from, the family I was raised in the time period I'm living in and the other individuals and experiences I've encountered during my life. \\
7 & Fortunately, the goals of deterrence of defense and of arms control are not always in conflict. For example, when we improve our command and control systems we improve our deterrent to aggression and at the same time we decrease the chance of a completely uncontrolled war. Should deterrence fail we have installed a number of both administrative and physical safeguards for our nuclear weapons which reduce as far as possible the chances of unauthorized use. The great emphasis we have placed on forces which can survive a nuclear attack from the Soviet not only serves to deter Soviet aggression but also greatly reduces the pressure on us to act precipitately in a crisis thus decreasing the danger of inadvertent or accidental war. \\
		\end{tabular}
}
	\caption{Examples of paragraphs exhibiting different levels of integrative complexity. The texts are taken from the training set and have been scored by certified professionals.}
	\label{tab:ic_example_text}
\end{table*}

\subsection{Limitations}

Our model is built using a training dataset of limited size, narrow range of genres and styles, and with scarce coverage of higher IC bands. The NLP methods used for feature extraction are general purpose and not optimized for the particular domain under study. This limitation does not prevent our method from obtaining good classification results but partially constrains the contribution that different types of syntactic features combined could add to the overall performance. To increase the number and variety of datapoints, we plan to collect more labels from trained human coders. In that respect, to ease this labelling process, our tool can be used to select candidate texts that are likely to belong to a variety of IC classes.

The external validation of our model is limited to one case-study of depression. Further replications of prior work should be undertaken to produce more evidence in support for or against the validity of any automatic system. These could use the original datasets, where possible, and also extend application to new datasets drawn from social media. For a better understanding of the potential and limits of our method, a deeper analysis on how automated IC scoring behaves when applied to different types of textual input---topic of discussion, length of text, cultural background of the authors---is also in order. The use of our tool to analyze Reddit should be considered just as a preliminary attempt to illustrate its potential. Further validation of the methodology is needed before it can be applied at scale and ``in the wild''. We therefore recommend to the researchers that may use our tool in the future to do it with caution. Most of all, the tool in its current stage should not be used for ethically-sensitive tasks such as decision- or policy-making.

Given these limitations, this work only scratches the surface of the studies that the method we propose could eventually enable. Since previous research has shown that Integrative Complexity is a good predictor of the richness of dialogue~\cite{conway2016conservatives}, we believe that IC has a role in tackling the resolution of conflicts in an increasingly polarized social media space.

\section*{Appendix}

Some examples of texts from the training set belonging to the 7 IC bands are reported in Table~\ref{tab:ic_example_text}.

\bibliographystyle{aaai}

\begin{thebibliography}{}
\small

\bibitem[\protect\citeauthoryear{Ambili and
  Rasheed}{2014}]{ambili2014automated}
Ambili, A.~K., and Rasheed, K.~M.
\newblock 2014.
\newblock Automated scoring of the level of integrative complexity from text
  using machine learning.
\newblock In {\em Proceedings of the 13th International Conference on Machine
  Learning and Applications (ICMLA)},  300--305.
\newblock IEEE.

\bibitem[\protect\citeauthoryear{Ashok, Feng, and
  Choi}{2013}]{ashok2013success}
Ashok, V.~G.; Feng, S.; and Choi, Y.
\newblock 2013.
\newblock Success with style: Using writing style to predict the success of
  novels.
\newblock In {\em Proceedings of the Conference on Empirical Methods in Natural
  Language Processing},  1753--1764.

\bibitem[\protect\citeauthoryear{Baker-Brown \bgroup et al\mbox.\egroup
  }{1990}]{baker1990coding}
Baker-Brown, G.; Ballard, E.~J.; Bluck, S.; De~Vries, B.; Suedfeld, P.; and
  Tetlock, P.~E.
\newblock 1990.
\newblock {\em Coding manual for conceptual/integrative complexity}.
\newblock University of British Columbia and University of California.

\bibitem[\protect\citeauthoryear{Boyd, Golder, and Lotan}{2010}]{boyd2010tweet}
Boyd, D.; Golder, S.; and Lotan, G.
\newblock 2010.
\newblock Tweet, tweet, retweet: Conversational aspects of retweeting on
  twitter.
\newblock In {\em Proceedings of the 43rd Hawaii International Conference on
  System Sciences (HICSS)},  1--10.
\newblock IEEE.

\bibitem[\protect\citeauthoryear{Budak and
  Agrawal}{2013}]{budak2013participation}
Budak, C., and Agrawal, R.
\newblock 2013.
\newblock On participation in group chats on twitter.
\newblock In {\em Proceedings of the 22nd International Conference on World
  Wide Web (WWW)},  165--176.
\newblock ACM.

\bibitem[\protect\citeauthoryear{Chen and Guestrin}{2016}]{xgboost}
Chen, T., and Guestrin, C.
\newblock 2016.
\newblock Xgboost: A scalable tree boosting system.
\newblock In {\em Proceedings of the 22nd International Conference on Knowledge
  Discovery and Data Mining (KDD)},  785--794.
\newblock ACM.

\bibitem[\protect\citeauthoryear{Cheng \bgroup et al\mbox.\egroup
  }{2017}]{cheng2017anyone}
Cheng, J.; Bernstein, M.; Danescu-Niculescu-Mizil, C.; and Leskovec, J.
\newblock 2017.
\newblock Anyone can become a troll: Causes of trolling behavior in online
  discussions.
\newblock In {\em Proceedings of the Conference on Computer Supported
  Cooperative Work and Social Computing (CSCW)},  1217--1230.
\newblock ACM.

\bibitem[\protect\citeauthoryear{Cheng, Danescu-Niculescu-Mizil, and
  Leskovec}{2015}]{cheng2015antisocial}
Cheng, J.; Danescu-Niculescu-Mizil, C.; and Leskovec, J.
\newblock 2015.
\newblock Antisocial behavior in online discussion communities.
\newblock In {\em Proceedings of the 9th International Conference on Weblogs
  and Social Media (ICWSM)}.
\newblock AAAI.

\bibitem[\protect\citeauthoryear{Conway \bgroup et al\mbox.\egroup
  }{2014}]{conway2014automated}
Conway, L.~G.; Conway, K.~R.; Gornick, L.~J.; and Houck, S.~C.
\newblock 2014.
\newblock Automated integrative complexity.
\newblock {\em Political Psychology} 35(5):603--624.

\bibitem[\protect\citeauthoryear{Conway \bgroup et al\mbox.\egroup
  }{2016}]{conway2016conservatives}
Conway, L.~G.; Gornick, L.~J.; Houck, S.~C.; Anderson, C.; Stockert, J.;
  Sessoms, D.; and McCue, K.
\newblock 2016.
\newblock Are conservatives really more simple-minded than liberals? the domain
  specificity of complex thinking.
\newblock {\em Political Psychology} 37(6):777--798.

\bibitem[\protect\citeauthoryear{Danescu-Niculescu-Mizil \bgroup et
  al\mbox.\egroup }{2012}]{danescu2012echoes}
Danescu-Niculescu-Mizil, C.; Lee, L.; Pang, B.; and Kleinberg, J.
\newblock 2012.
\newblock Echoes of power: Language effects and power differences in social
  interaction.
\newblock In {\em Proceedings of the 21st international conference on World
  Wide Web},  699--708.
\newblock ACM.

\bibitem[\protect\citeauthoryear{De~Choudhury and De}{2014}]{de2014mental}
De~Choudhury, M., and De, S.
\newblock 2014.
\newblock Mental health discourse on reddit: Self-disclosure, social support,
  and anonymity.
\newblock In {\em Proceedings of the 8th International Conference on Weblogs
  and Social Media (ICWSM)}.
\newblock AAAI.

\bibitem[\protect\citeauthoryear{De~Choudhury \bgroup et al\mbox.\egroup
  }{2009}]{de2009makes}
De~Choudhury, M.; Sundaram, H.; John, A.; and Seligmann, D.~D.
\newblock 2009.
\newblock What makes conversations interesting?: themes, participants and
  consequences of conversations in online social media.
\newblock In {\em Proceedings of the 18th International Conference on World
  Wide Web (WWW)},  331--340.
\newblock ACM.

\bibitem[\protect\citeauthoryear{Gilbert}{2014}]{gilbert2014vader}
Gilbert, C. H.~E.
\newblock 2014.
\newblock Vader: A parsimonious rule-based model for sentiment analysis of
  social media text.
\newblock In {\em Proceedings of the 8th International Conference on Weblogs
  and Social Media (ICWSM)}.
\newblock AAAI.

\bibitem[\protect\citeauthoryear{Gosling, Rentfrow, and
  Swann~Jr}{2003}]{gosling2003very}
Gosling, S.~D.; Rentfrow, P.~J.; and Swann~Jr, W.~B.
\newblock 2003.
\newblock A very brief measure of the big-five personality domains.
\newblock {\em Journal of Research in personality} 37(6):504--528.

\bibitem[\protect\citeauthoryear{Hofmann, Wickham, and
  Kafadar}{2017}]{hofmann2017value}
Hofmann, H.; Wickham, H.; and Kafadar, K.
\newblock 2017.
\newblock Value plots: Boxplots for large data.
\newblock {\em Journal of Computational and Graphical Statistics}
  26(3):469--477.

\bibitem[\protect\citeauthoryear{Kim, Bak, and Oh}{2012}]{kim2012you}
Kim, S.; Bak, J.; and Oh, A.~H.
\newblock 2012.
\newblock Do you feel what i feel? social aspects of emotions in twitter
  conversations.
\newblock In {\em Proceedings of the 6th International Conference on Weblogs
  and Social Media (ICWSM)}.
\newblock AAAI.

\bibitem[\protect\citeauthoryear{Li, Bandar, and McLean}{2003}]{li2003approach}
Li, Y.; Bandar, Z.~A.; and McLean, D.
\newblock 2003.
\newblock An approach for measuring semantic similarity between words using
  multiple information sources.
\newblock {\em IEEE Transactions on knowledge and data engineering}
  15(4):871--882.

\bibitem[\protect\citeauthoryear{McCullough and Conway~III}{2018a}]{oscars}
McCullough, H., and Conway~III, L.~G.
\newblock 2018a.
\newblock ``and the oscar goes to...'': Integrative complexity's predictive
  power in the film industry.
\newblock {\em Psychology of Aesthetics, Creativity, and the Arts} 12(4):392.

\bibitem[\protect\citeauthoryear{McCullough and Conway~III}{2018b}]{piggy}
McCullough, H., and Conway~III, L.~G.
\newblock 2018b.
\newblock The cognitive complexity of miss piggy and osama bin laden: Examining
  linguistic differences between fiction and reality.
\newblock {\em Psychology of Popular Media Culture} 7(4):518.

\bibitem[\protect\citeauthoryear{Miller}{1995}]{wordnet}
Miller, G.~A.
\newblock 1995.
\newblock Wordnet: A lexical database for english.
\newblock {\em Communications of the ACM} 38(11):39--41.

\bibitem[\protect\citeauthoryear{Niculae and
  Danescu-Niculescu-Mizil}{2016}]{niculae2016conversational}
Niculae, V., and Danescu-Niculescu-Mizil, C.
\newblock 2016.
\newblock Conversational markers of constructive discussions.
\newblock In {\em Proceedings of the Conference of the North American
  Association for Computational Linguistics (NAACL-HLT)}.

\bibitem[\protect\citeauthoryear{Pennebaker, Francis, and Booth}{2001}]{liwc}
Pennebaker, J.~W.; Francis, M.~E.; and Booth, R.~J.
\newblock 2001.
\newblock Linguistic inquiry and word count.
\newblock {\em Mahway: Lawrence Erlbaum Associates} 71.

\bibitem[\protect\citeauthoryear{Purohit \bgroup et al\mbox.\egroup
  }{2014}]{purohit2014understanding}
Purohit, H.; Ruan, Y.; Fuhry, D.; Parthasarathy, S.; and Sheth, A.~P.
\newblock 2014.
\newblock On understanding the divergence of online social group discussion.
\newblock In {\em Proceedings of the 8th International Conference on Weblogs
  and Social Media (ICWSM)}.
\newblock AAAI.

\bibitem[\protect\citeauthoryear{Speer, Chin, and Havasi}{2017}]{conceptnet}
Speer, R.; Chin, J.; and Havasi, C.
\newblock 2017.
\newblock Conceptnet 5.5: An open multilingual graph of general knowledge.
\newblock In {\em Thirty-First AAAI Conference on Artificial Intelligence},
  4444--4451.

\bibitem[\protect\citeauthoryear{Suedfeld and
  Bluck}{1993}]{suedfeld1993changes}
Suedfeld, P., and Bluck, S.
\newblock 1993.
\newblock Changes in integrative complexity accompanying significant life
  events: Historical evidence.
\newblock {\em Journal of personality and Social Psychology} 64(1):124.

\bibitem[\protect\citeauthoryear{Suedfeld and
  Tetlock}{1977}]{suedfeld1977integrative}
Suedfeld, P., and Tetlock, P.
\newblock 1977.
\newblock Integrative complexity of communications in international crises.
\newblock {\em Journal of Conflict Resolution} 21(1):169--184.

\bibitem[\protect\citeauthoryear{Suedfeld, Tetlock, and
  Ramirez}{1977}]{suedfeld1977war}
Suedfeld, P.; Tetlock, P.~E.; and Ramirez, C.
\newblock 1977.
\newblock War, peace, and integrative complexity: Un speeches on the middle
  east problem, 1947--1976.
\newblock {\em Journal of Conflict Resolution} 21(3):427--442.

\bibitem[\protect\citeauthoryear{Suedfeld, Tetlock, and
  Streufert}{1992}]{sts1992}
Suedfeld, P.; Tetlock, P.~E.; and Streufert, S.
\newblock 1992.
\newblock Conceptual/integrative complexity.
\newblock In Smith, C.~P., ed., {\em Motivation and personality: Handbook of
  thematic content analysis}. Cambridge University Press.
\newblock  393--400.

\bibitem[\protect\citeauthoryear{Tausczik and
  Pennebaker}{2010}]{tausczik2010psychological}
Tausczik, Y.~R., and Pennebaker, J.~W.
\newblock 2010.
\newblock The psychological meaning of words: Liwc and computerized text
  analysis methods.
\newblock {\em Journal of Language and Social Psychology} 29(1):24--54.

\bibitem[\protect\citeauthoryear{Tchokni, S{\'e}aghdha, and
  Quercia}{2014}]{tchokni2014emoticons}
Tchokni, S.~E.; S{\'e}aghdha, D.~O.; and Quercia, D.
\newblock 2014.
\newblock Emoticons and phrases: Status symbols in social media.
\newblock In {\em Proceedings of the 8th International Conference on Weblogs
  and Social Media (ICWSM)}.
\newblock AAAI.

\bibitem[\protect\citeauthoryear{Thoemmes and
  Conway}{2007}]{thoemmes2007integrative}
Thoemmes, F.~J., and Conway, L.~G.
\newblock 2007.
\newblock Integrative complexity of 41 us presidents.
\newblock {\em Political Psychology} 28(2):193--226.

\bibitem[\protect\citeauthoryear{Vosoughi and Roy}{2016}]{vosoughi2016tweet}
Vosoughi, S., and Roy, D.
\newblock 2016.
\newblock Tweet acts: A speech act classifier for twitter.
\newblock In {\em Proceedings of the 10th International Conference on Weblogs
  and Social Media (ICWSM)}.
\newblock AAAI.

\bibitem[\protect\citeauthoryear{Winter}{1993}]{winter1993slot}
Winter, D.~A.
\newblock 1993.
\newblock Slot rattling from law enforcement to lawbreaking: A personal
  construct theory exploration of police stress.
\newblock {\em International Journal of Personal Construct Psychology}
  6(3):253--267.

\bibitem[\protect\citeauthoryear{Zhang \bgroup et al\mbox.\egroup
  }{2018}]{zhang18conversations}
Zhang, J.; Chang, J.; Danescu-Niculescu-Mizil, C.; Dixon, L.; Hua, Y.;
  Taraborelli, D.; and Thain, N.
\newblock 2018.
\newblock Conversations gone awry: Detecting early signs of conversational
  failure.
\newblock In {\em Proceedings of the 56th Meeting of the Association for
  Computational Linguistics},  1350--1361.
\newblock ACL.

\end{thebibliography}

\end{document}